\documentclass[acmsmall,natbib,screen=true,]{acmart} 

\usepackage{booktabs} 
\usepackage[skip=0pt]{caption}
\usepackage{multicol}
\usepackage{multirow}
\usepackage[inline]{enumitem}
\usepackage{bbm}
\usepackage{acronym}
\usepackage{algorithm}
\usepackage{algorithmic}
\usepackage{array}
\usepackage{subfigure}

\AtBeginDocument{%
  \providecommand\BibTeX{{%
    \normalfont B\kern-0.5em{\scshape i\kern-0.25em b}\kern-0.8em\TeX}}}
\newcommand\BibTeX{B\textsc{ib}\TeX}
\acrodef{NLP}{natural language processing}
\acrodef{MIFN}{mixed information flow network}
\acrodef{BPTT}{back-propogation through time}
\acrodef{CDSR}{cross-domain sequential recommendation}
\acrodef{CF}{collaborative filtering}
\acrodef{DSSM}{deep structured semantic model}
\acrodef{GRU}{gated recurrent unit}
\acrodef{KG}{knowledge graph}
\acrodef{KNN}{K-Nearest neighbors}
\acrodef{MC}{Markov chains}
\acrodef{MDP}{Markov decision process}
\acrodef{MF}{matrix factorization}
\acrodef{MLP}{multilayer perceptron}
\acrodef{RBM}{restricted Boltzmann machines}
\acrodef{ReLU}{rectified linear unit}
\acrodef{RNN}{recurrent neural network}
\acrodef{RS}{recommender system}
\acrodef{SR}{sequential recommendation}
\acrodef{BTU}{behavior transfer unit}
\acrodef{KTU}{knowledge transfer unit}
\acrodef{CDGCM}{cross-domain graph convolutional mechanism}
\acrodef{GCN}{graph convolutional network}

\setcopyright{acmcopyright}
\copyrightyear{2020}
\acmYear{2020}
\acmDOI{xx.xxxx/xxxxxxx.xxxxxxx}
\acmPrice{15.00}
\acmISBN{978-1-4503-9999-9/18/06}
\acmJournal{TKDD}

\author{Muyang Ma}
\orcid{}
\affiliation{%
\institution{Shandong University}
\city{Jinan}
\country{China}
}
\email{muyang0331@gmail.com}

\author{Pengjie Ren}
\orcid{}
\affiliation{%
\institution{Shandong University}
\city{Jinan}
\country{China}
}
\email{jay.ren@outlook.com}

\author{Zhumin Chen}
\orcid{}
\affiliation{%
\institution{Shandong University}
\city{Jinan}
\country{China}
}
\email{chenzhumin@sdu.edu.cn}

\author{Zhaochun Ren}
\orcid{}
\affiliation{%
\institution{Shandong University}
\city{Qingdao}
\country{China}
}
\email{zhaochun.ren@sdu.edu.cn}

\author{Lifan Zhao}
\orcid{}
\affiliation{%
\institution{Shandong University}
\city{Qingdao}
\country{China}
}
\email{mogician233@outlook.com}

\author{Jun Ma}
\orcid{}
\affiliation{%
\institution{Shandong University}
\city{Jinan}
\country{China}
}
\email{majun@sdu.edu.cn}

\author{Maarten de Rijke}
\orcid{0000-0002-1086-0202}
\affiliation{
 \institution{University of Amsterdam and Ahold Delhaize}
 \city{Amsterdam and Zaandam}
 \country{The Netherlands}
}
\email{m.derijke@uva.nl}

\thanks{Pengjie Ren and Zhumin Chen are the corresponding authors.}

\begin{document}
 
\title[MIFN: Mixed Information Flow Network]{Mixed Information Flow for Cross-domain Sequential Recommendations}

\begin{abstract}
Cross-domain sequential recommendation is the task of predict the next item that the user is most likely to interact with based on past sequential behavior from multiple domains.
One of the key challenges in \acl{CDSR} is to grasp and transfer the flow of information from multiple domains so as to promote recommendations in all domains.
Previous studies have investigated the flow of \emph{behavioral} information by exploring the connection between items from different domains.
The flow of knowledge (i.e., the connection between knowledge from different domains) has so far been neglected.
In this paper, we propose a \acl{MIFN} for \acl{CDSR} to consider both the flow of behavioral information and the flow of knowledge by incorporating a \acl{BTU} and a \acl{KTU}.
The proposed \acl{MIFN} is able to decide when cross-domain information should be used and, if so, which cross-domain information should be used to enrich the sequence representation according to users' current preferences.
Extensive experiments conducted on four e-commerce datasets demonstrate that \acl{MIFN} is able to further improve recommendation performance in different domains by modeling mixed information flow.
\end{abstract}

\ccsdesc[500]{Information systems~Recommender systems}

\keywords{Cross-domain recommendation, Sequential recommendation, Knowledge base, Graph transfer}

\maketitle

\acresetall


\section{Introduction}
\begin{figure}[ht]
\centering
\includegraphics[clip,trim=7mm 7mm 7mm 9mm, width=0.8\columnwidth]{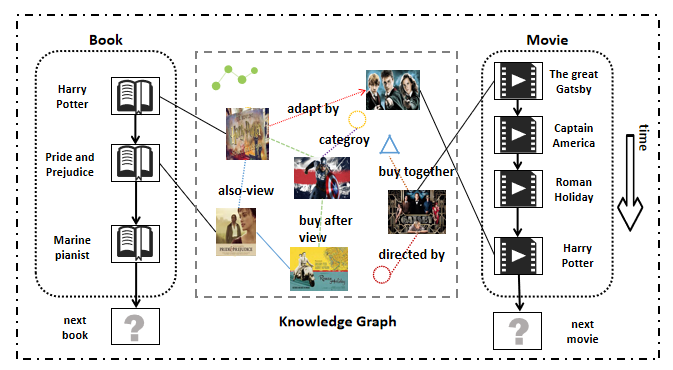}
\vspace*{2mm}
\caption{Illustration of a connection between a user's behavioral information across domains via a \acl{KG}. Lines in different colors represent different connections.}
\label{story}
\end{figure}

Sequential recommendation (SR\acused{SR}) aims to predict the next item that the user is most likely to interact with based on her/his past sequential behavior (e.g., clicks on items)~\cite{HRNN}.
Recently, \acf{CDSR} has emerged as a way to promote recommendation performance by leveraging and combining information from different domains~\cite{ma2019pi}.
Users usually have related preferences in different domains, such as finding a movie with a certain style or looking for a book written by a well-known author, as illustrated in Figure~\ref{story}.
One of the key challenges in \ac{CDSR} is to capture and transfer useful information about related preferences across different domains.

\noindent%
\citet{cross-domain_novelty-seeking_sr} and \citet{ma2019pi} have shown that behavioral information across domains is helpful for improving recommendation performance.
However, behavioral information by itself can only support the use of cross-domain connections in a limited manner. 
Behavioral information may be insufficient for a model to capture fine-grained connections between item attributes or features.
For example, as illustrated in Figure~\ref{story}, assume there is a user who has read \textit{Harry Potter} (the book) and watched \textit{Captain America} (the movie).
If there is no external knowledge to indicate that both items belong to the category of ``fantasy,'' it is difficult for the model to capture this connection based solely on the user's behavior from both domains.
We hypothesize that enabling a flow of knowledge across different domains is able to alleviate this issue.
As a result, for a user who has read the book \textit{The Great Gatsby}, we then can recommend her/him a movie having the same name or movies featuring the same category of ``tragic love,'' such as \textit{Atonement}, \textit{Waterloo Bridge}, and so on, when she/he logs on to the movie recommendation system.

There is a growing body of work aimed at improving recommendation performance by using knowledge~\cite{10.1145/3336191.3371884,lin2019explainable}.
Of particular relevance to us is work that has proposed to incorporate knowledge and combine it with behavioral information for \ac{SR}~\cite[see, e.g.,][]{KSR,SRGNN,huang2019taxonomy}.
However, this work targets a single domain recommendation scenario.
The situation is dramatically different in cross-domain scenarios where it is necessary to distinguish information from different domains and effectively link them.
We need to select behavioral and knowledge related to users' current preference, and determine when and what to use in order to learn a better sequence representation.

To address the issue of using behavioral information and knowledge across domains, we propose a \acfi{MIFN} to consider mixed information flow across domains, i.e., the flow of behavioral information as well as the flow of knowledge.
The former is based on user's behavior, which captures the temporal connection between the items they have interacted with, while the latter takes cross-domain knowledge as a bridge to connect different domains to obtain better cross-domain sequence representations.
First, we employ a \acfi{BTU} to grasp useful information from the flow of behavioral information, which can extract information related to the user's preference and then transfer it to another domain at the level of user behavior level.
Then, we propose a \acfi{KTU} that is guided by the user's preference to model the connection between items from different domains; we introduce a \acl{CDGCM} to distinguish items in the \ac{KG} and grasp useful information for fusion at the knowledge level.
Finally, we generate recommendations based on the fusion of the two types of information.
During learning, \ac{MIFN} is jointly trained on multiple domains in an end-to-end back-propagation training paradigm.
Experiments on the Amazon datasets show that \ac{MIFN} outperforms state-of-the-art methods in terms of MRR and Recall.

To sum up, the contributions of this work are as follows:
\begin{itemize}[leftmargin=*,nosep]
\item We propose a mixed information flow framework, \ac{MIFN}, for \ac{CDSR}, which consists of a \acl{BTU} and a \acl{KTU} to simultaneously model the flow of behavioral information and of knowledge across domains.
\item We devise a \acl{CDGCM} to disseminate item information in the \ac{KG}, which leads to the better up-to-date item representation.
\item We conduct experiments to demonstrate that \ac{MIFN} is able to improve recommendation performance in different domains by modeling mixed information flow.
\end{itemize}


\section{Related work}
In this section, we briefly introduce related work from the following categories: 
\begin{enumerate*}
\item sequential recommendation, 
\item cross-domain recommendation, and 
\item knowledge-aware recommendation.
\end{enumerate*}

\subsection{Sequential recommendations}
Early work on \acp{RS} typically use \ac{CF} to generate recommendations \cite{huCF2018tkdd} according to users' preferences reflected in similar items such as \ac{KNN} or \ac{MF} algorithms.
Such methods do not consider sequential aspects.
More recently, however, \ac{SR} and next-basket recommendation have witnessed rapid developments.

Before the widespread application of deep learning, \acp{MC} \cite{zimdars2001using,rendle2010factorizing,chen2012playlist,he2016fusing} and \acp{MDP} \cite{shani2005mdp,wu2013personalized} have been used to predict users’ next action given information about their past behavior \cite{wang2015learning,yap2012effective}.
All these methods take into account the sequential characteristics.
However, there are considerable challenges with the size of the state when considering the entire sequence \cite{quadrana2018sequence}.

\Acp{RNN} have been introduced to \ac{SR} to handle variable-length sequential data.
\citet{GRU4rec} are the first to leverage \acp{RNN} for \ac{SR}.
They utilize session-parallel mini-batch training and employ ranking-based loss functions to train the model.
Then, \citet{improveGRU4rec} propose two techniques to improve the performance, i.e., data augmentation and a method to account for shifts in the input data distribution.
\citet{NARM} incorporate an attention mechanism into the encoder to capture the users' main preference in the current sequence.
\citet{repeatnet} point out the repeat consumption in \ac{SR}, where the same item is re-consumed repeatedly over time. 
\citet{HRNN} propose a hierarchical \ac{RNN} model that can relay and evolve latent hidden states of the \acp{RNN} across user sequences.
\citet{donkers2017sequential} explicitly model user information in a gated architecture with extra input layers for \ac{GRU}.
Memory enhanced \ac{RNN} has been well studied for \ac{SR} recently.
\citet{chen2018sequential_memNet} introduce a memory mechanism to \ac{SR} and design a memory-augmented neural network integrated with the insights of \ac{CF}.
\citet{wang2019collaborative} propose two parallel memory modules: one to model a user’s own information in the current sequence and the other to exploit collaborative information in neighborhood sequences.
\citet{SRGNN} argue that prior work on conventional sequential methods neglects complex transitions between items.
They model the sequence as graph-structured data and then represent it as the composition of global preference and the current preference of that sequence using an attention network.
\citet{zhang2019feature} propose a feature-level deeper self-attention network to capture transition patterns between features of items by integrating various heterogeneous features.
\citet{bert_sr_2019} argue that previous work often assumes a rigidly ordered sequence, which is not always practical.
They employ deep bidirectional self-attention to model a user's behavioral sequences.

In addition to sequential information, auxiliary information is also vital for \ac{SR}.
\citet{hidasi2016parallel} investigate how to add item property information such as text and images to an \acp{RNN} framework and introduce a number of parallel \ac{RNN} (p-\ac{RNN}) architectures.
\citet{liu2016context} incorporate contextual information into \ac{SR} and propose a context-aware \ac{RNN} model to capture external situations and lengths of time intervals.
\citet{dwelltimernn} explore a user's dwell time based on an existing \ac{RNN}-based framework by boosting items above a predefined dwell time threshold. 
\citet{ma2018mention} propose a cross-attention memory network for multi-modal tweets via both textual and visual information.
\citet{review_driven_sr2019} study how to enlist the semantic signals covered by user reviews for the task of \ac{CF}.
They propose a neural review-driven model by considering users’ intrinsic preference and sequential patterns.
To investigate the influence of temporal sentiments on user preference, \citet{sentiSR2020} propose to generate preferences by guiding user behavior through sequential sentiments.
They design a dual-channel fusion mechanism to match and guide sequential user behavior, and to assist in preference generation.
\citet{masr2020} model the effect of context information on \ac{SR} and train the model in an adversarial manner by proposing multiple context-specific discriminators to evaluate the generated sub-sequence from the perspectives of different contexts.

Although these studies have made great progress, none of them has considered how to combine knowledge information under cross-domain situations.

\subsection{Cross-domain recommendations}
Cross-domain recommendation has emerged as a potential solution to the cold-start and data-sparse problem \cite{enhance-person,transfer-cf} in \ac{RS}.
It aims to mitigate the lack of data by exploiting user preference and item attributes in domains distinct but related to the target domain \cite{fernandez2019addressing}.

Traditional methods for cross-domain recommendation can be grouped into two main categories \cite{cd_survey}.
One category of methods \textit{aggregates information} across different domains.
According to different aggregating strategies, such methods can be further divided into three groups~\cite{fernandez2019addressing}.
The first group is \textit{merging user preference} (e.g., ratings, transaction behavior, and browsing logs) from different domains to obtain better a preference representation so as to improve the recommendation performance in the target domain.
The merge operation is performed by merging a multi-domain rating matrix \cite{BerkovskyKR07,SahebiB13}, leveraging users' social influence \cite{crosssys-usermodel,Fernandez-TobiasCP13}, linking users' preference by a multi-domain graph \cite{CremonesiTT11,TiroshiBKCK13} or user behavioral information features \cite{LoniSLH14,MaZWLLM18}.
The second group is \textit{mediating user modeling data} in the source domain to explore the connection between users or items so as to make recommendations in the target domain especially for cold start users.
For example, \citet{TiroshiK12} and \citet{ShapiraRF13} propose to find similar neighbors and transfer user-user similarity to the target domain.
The third group is \textit{combining single-domain recommendations} (e.g., rating matrices, probability distributions), in which recommendations are generated independently for each domain and later aggregated for the final recommendation.
In contrast to the second group, this type of aggregation strategy aims to model the weights assigned to recommendations coming from different domains.
For example, \citet{Givon2009} focus on book recommendations accomplished by a \ac{CF} method and model-based recommendations, relying on the similarity of a book and the user’s model, as well as the book content and tags.
And the final recommendations are combined in a weighted manner.
The other category of cross-domain recommendation aims to \textit{transfer information} from the source domain to the target domain by means of shared latent features or rating patterns.
\citet{hu2013} propose tensor-based factorization to share latent features between different domains by using the same parameters in both factorization models.
\citet{cd-cf} propose a code-book-transfer by co-clustering the source domain rating matrix and exploit it in the target domain to transfer rating patterns across different domains.
Similarly, \citet{improving2015tkdd} focus on extending clustering-based \ac{MF} in a single domain into multiple domains through overlapping users.
%

In order to model more complex connections across different domains, a variety of deep learning methods have been proposed for cross-domain recommendation.
\citet{DSSM2015} propose a multi-view deep learning recommendation system by using rich auxiliary features to represent users from different domains.
Then, \citet{CCCFNet} propose a multi-view neural framework of a dual network for user and item, each network models \ac{CF} information (user and item embeddings) and content information (user preference for item features), which ties \ac{CF} and content-based filtering together.
\citet{Conet} propose a model using a cross-stitch network \cite{misra2016cross} to learn complex user behavioral information based on neural \ac{CF}~\cite{ncf_he}. 
\citet{itemsilkroad2017} propose to combine user behavioral information in information domains and user-user connection in social domains to do recommendation.
%
%
\citet{wang2019solving} embed item-level information and cluster-level correlative information from different domains into a unified framework.
\citet{gao2019cross} transfer item embeddings across domains without sharing user-relevant data.
\citet{li2019ddtcdr} develop a latent orthogonal mapping method to extract user preference over multiple domains while preserving connection between users across different latent spaces based on the mechanism of dual learning.
\citet{krishnan2020transfer} propose to guide neural \ac{CF} with domain-invariant components shared across the dense and sparse domains, improving  user and item representations learned in the sparse domains. 
They leverage contextual invariances across domains to develop these shared modules.
\citet{zhao2020catn} propose to model user preference transfer at the aspect-level derived from reviews, which does not require overlapping users or items in all domains.
\citet{ye2020dcdir} utilize cross-domain mechanism to promote recommendations for cold start users in insurance domain.
They design a meta-path based method over complex insurance products to learn better item representations and learn the mapping function between domains through the overlapping users.
Despite the fact that the listed methods above have been proven to be effective, they cannot be directly applied to \acp{SR}.

Recently, cross-domain recommendation has been introduced to \acp{SR} as well.
\citet{cross-domain_novelty-seeking_sr} propose a novelty seeking model based on sequences in multi-domains to model an individual’s propensity by transferring novelty seeking traits learned from a source domain for improving the accuracy of recommendations in the target domain.
\citet{ma2019pi} study \ac{CDSR} in a shared-account scenario.
They propose a novel gating mechanism to extract and share user-specific information between domains.

%
Although some studies have begun to explore \ac{CDSR}, they only focus on user behavioral information to conduct information transfer, and neglect to explore extra knowledge to promote sequence representation across domains.

\subsection{Knowledge-aware recommendations}
Considerable efforts have been made to utilize side-information, especially \aclp{KG}, to enhance the performance of recommendations.
\citet{zhao2016tkdd} propose a graph-based method to iteratively update user and item distributions in a heterogeneous user-item graph and incorporate them as features into the \ac{MF} for item recommendations.
\citet{zhang2016collaborative} combine \ac{CF} with structural knowledge, textual knowledge and visual knowledge in a unified framework.
\citet{ai2018learning} apply TransE on the graph including users, items and their connections, which casts the recommendation task as a plausibility prediction task.
As \acp{GCN} have been shown to be effective on many tasks~\cite{kipf2016semi}, there have been a number of publications that propose variants of \acp{GCN} for recommendation by considering different types of information.
\citet{RippleNet} simulate users' hierarchical preferences over knowledge entities by extending users' potential preferences along links in a \ac{KG}.
\citet{KGCN} consider the connections among items based on higher-order entity features in \acp{KG}. 
\citet{wang2019kgat} explicitly model the high-order connections in \acp{KG} by employing an attention mechanism to discriminate the importance of the neighbors.
\citet{ma2019jointly} propose a joint framework to integrate the induction of explainable rules from \acp{KG} with the construction of a rule-guided recommendation model.
\citet{xian2019reinforcement} perform explicit reasoning with knowledge so that the recommendations are supported by an interpretable inference procedure via a policy-guided reinforcement learning approach.

Not surprisingly, \acp{KG} has also been considered in \acp{SR}.
\citet{KSR} are the first to integrate \acp{KG} into \ac{SR}; they utilize \acp{RNN} to capture user sequential preference and knowledge memory networks to capture attribute-level user preference.
\citet{song2019session} model users' social influence with a graph-attention neural network, which dynamically infers the influencers based on the users' current preference.
\citet{huang2019taxonomy} introduce a taxonomy-aware memory-based multi-hop reasoning architecture by incorporating taxonomy data as structural knowledge to enhance the reasoning capacity.
%

However, no previous work has considered \acp{KG} for \ac{SR} in a cross-domain scenario, which brings new challenges, e.g., how to find useful and accurate cross-domain knowledge to improve information transfer across domains to promote the performance in both domains.


\section{Method}
In this section, we first give a formulation of the \acs{CDSR} task.
Then, we give an overview of our model \ac{MIFN}.
Finally, we describe each component of \ac{MIFN} in detail.
Table~\ref{symbols} summarizes the main symbols and notation used in this paper.

\begin{table*}[]
\caption{Summary of main symbols and notation used in the paper.}
\label{symbols}
\begin{tabular}{lm{12cm}}
\toprule

$\mathbbm{A}$    & Item set for domain $A$. \\
$\mathbbm{B}$    & Item set for domain $B$. \\
$n$    & The number of item set for domain $A$, i.e., $n = |\mathbbm{A}|$. \\
$m$    & The number of item set for domain $B$, i.e., $m = |\mathbbm{B}|$. \\
$A_i$    & The interacted item at time step $i$ from domain $A$. \\
$B_j$    & The interacted item at time step $j$ from domain $B$. \\
$S$    & Hybrid interaction sequence, i.e., $S=[A_1$, $B_1$, $B_2$, \ldots, $A_i$, \ldots, $B_j$, $\ldots]$. \\
$\mathbb{S}$    & Set of all hybrid interaction sequences in the training set. \\
$S_A$    & A sub-sequence of $S$ which only contains items from domain $A$. \\
$S_B$    & A sub-sequence of $S$ which only contains items from domain $B$. \\
$e_{A_k} \in E_A$    & $e_{A_k}$ represents any entity in the \ac{KG} from domain $A$; $E_A$ is the set of all entities from domain $A$. \\
$e_{B_k} \in E_B$    & $e_{B_k}$ represents any entity in the \ac{KG} from domain $B$; $E_B$ is the set of all entities from domain $B$. \\
$e_k \in E$    & $e_k$ represents any entity in the \ac{KG}; $E$ is the set of all entities; note that  $E = E_A \cup E_B$.\\
$R$    & Relation set in the \ac{KG}. \\
$h_{A_i} \in H_A$    & $h_{A_i}$ represents the item representation of item $A_i$; $H_A$ is the set of all item representations for $S_A$. \\
$h_{B_j} \in H_B$    & $h_{B_j}$ represents the item representation of item $B_j$; $H_B$ is the set of all item representations for $S_B$. \\
$h_{(A \rightarrow B)_i}$    & Transferred behavioral information flow from domain $A$ to domain $B$ at time step $i$. \\
$N_i(k)$    & Neighbor entity set of entity $e_k$ from the same domain as $e_k$. \\
$N_c(k)$    & Neighbor entity set of entity $e_k$ from the complementary domain. \\


\bottomrule
\end{tabular}
\end{table*}

\subsection{Task formulation}
\Acf{CDSR} aims to predict the next item the user is mostly likely to interact with in multiple domains simultaneously, by mining users' previous sequential behavior over a period of time.
In this work, we take two domains (i.e., domain $A$ and $B$) as an example, e.g., watching movies, reading books.
Let $\mathbbm{A}=\{A_1$, $A_2$, $A_3$, \ldots, $A_n\}$ denote the item set for domain $A$, which consist of $n$ unique items.
Similarly, let $\mathbbm{B}=\{B_1$, $B_2$, $B_3$, \ldots, $B_m\}$ denote the item set for domain $B$, which consist of $m$ unique items.
A \textit{hybrid interaction sequence} from the two domains $A$ and $B$ has the form $S=[A_1$, $B_1$, $B_2$, \ldots, $A_i$, \ldots, $B_j$, $\ldots]$, 
where $A_i \in \mathbbm{A} $ $(1 \leq  i \leq n)$ and $B_j \in \mathbbm{B} $ $(1 \leq  j \leq m)$ are the indices of consumed items in domain $A$ and $B$, respectively.

We also associate each $S$ with a \acf{KG}, which is defined over an entity set $E$ and a relation set $R$, containing a set of \ac{KG} triples.
A triple $\langle e_1, r, e_2 \rangle$ represents a relation $r$ $\in$ $R$ between two entities $e_1$ and $e_2$ from $E$.
In the cross-domain scenario, entities in the \ac{KG} come from different domains, hence we represent them as $E_A$ and $E_B$.
For example, the triple \textit{$\langle e_{A_1}, Is\_the\_same\_category, e_{B_3} \rangle$} means that entity $e_{A_1}$ from domain $A$ has the same category as entity $e_{B_3}$ from domain $B$.
Since we aim to link recommended items to \ac{KG} entities, an item set can be considered as a subset of \ac{KG} entity set, i.e., $\mathbbm{A} \subseteq E_A$ and $\mathbbm{B} \subseteq E_B$.
When extracting \ac{KG} information for each sequence $S$, we also refer to the ``items'' in $S$ as ``item entities''.
We will explain the details of the \ac{KG} construction method in Section~\ref{graph_constructor}.

Based on these preliminaries, we are ready to define the \ac{CDSR} task.
Formally, given $S$ and $\langle E, R \rangle$, we formulate \acs{CDSR} as a task of evaluating the recommendation probabilities for all candidates in both domains respectively, as shown in Eq. ~\ref{general_equations}:
\begin{equation}
\label{general_equations}
\begin{split}
P(A_{i+1}|S, \langle E, R \rangle) & \sim f_A(S, \langle E, R \rangle) \\
P(B_{j+1}|S, \langle E, R \rangle) & \sim f_B(S, \langle E, R \rangle), \\
\end{split}
\end{equation}
where $P(A_{i+1}|S, \langle E, R \rangle)$ denotes the probability of recommending the next item $A_{i+1}$ in domain $A$ given the hybrid interaction sequence $S$ and \ac{KG} $\langle E, R \rangle$.
$f_A(S, \langle E, R \rangle)$ is the model or function used to estimate $P(A_{i+1}|S, \langle E, R \rangle)$.
Similar definitions apply to $P(B_{j+1}|S, \langle E, R \rangle)$ and $f_B(S, \langle E, R \rangle)$.

\subsection{Knowledge graph construction}
\label{graph_constructor}
In this work, we extract data (entities and relations) from the Amazon product collection\footnote{\url{https://jmcauley.ucsd.edu/data/amazon}} as the complete \ac{KG}, which is collected from massive user logs.
Besides, we also crawl some relations between the entities from Wikipedia.\footnote{\url{https://en.wikipedia.org/}}
The entities include ``movies,'' ``books,'' ``kitchenware,'' and ``food,'' each of which corresponds to one domain.
The relations include ``\textit{Also\_buy},'' ``\textit{Also\_view},'' ``\textit{Buy\_together},'' ``\textit{Buy\_after\_viewing},'' ``\textit{Adapted\_from},'' and ``\textit{Is\_the\_same\_category}.''
For example, $\langle$steak (the food), Buy\_together, saucepan (the kitchenware)$\rangle$ means that the user also buys the saucepan while buying the steak.
However, the complete \ac{KG} contains a large number of related entities, and complicated relations among these entities, which will raise memory and computational efficiency issues.
Therefore, we propose to extract a \ac{KG} from the complete \ac{KG} for each hybrid interaction sequence $S$.
We require that, given any pair of items from both domains respectively, we can find at least one path in the \ac{KG} that connects them.

\begin{algorithm}[htb]
\caption{\ac{KG} construction algorithm for each hybrid interaction sequence.}
\label{alg:Extractgraph}
\begin{algorithmic}[1]
\REQUIRE ~~\\ 
Hybrid interaction sequence, $S$;\\
Complete \ac{KG} triples, $\langle E, R, E \rangle$;\\
Maximum hop count, $H$;\\
Number of entities in the \ac{KG}, $N$;\\
\ENSURE ~~\\ 
Multiple relational adjacency matrix, $A$;
\STATE Divide the hybrid interaction sequence $S$ into $S_A$ and $S_B$;
\label{ code:fram:extract }
\STATE set $\vartheta^0_1$ = $S_A$, $\vartheta^0_2$ = $S_B$, $\mathit{connected}=\mathit{false}$;
\FOR{$k \in range(H)$}
\STATE Extract related triples $\vartheta^k_1=\{(h,r,t)|(h, r, t)\in \langle E, R, E \rangle$ and $h \in \vartheta^{k-1}_1 \}$;
\STATE Extract related triples $\vartheta^k_2=\{(h,r,t)|(h, r, t)\in \langle E, R, E \rangle$ and $h \in \vartheta^{k-1}_2 \}$;
\STATE $\mathit{connected}$ = Is\_Connect($\vartheta^k_1$, $\vartheta^k_2$)
\IF{$\mathit{connected}=\mathit{true}$ or $k$=$H$-1}
\STATE $\theta$ = Select\_triples($\vartheta^k_1$, $\vartheta^k_2$, $S_A$, $S_B$);
\STATE $A$ = Construct\_Adjacency\_matrix($\theta$);
\STATE break;
\ENDIF
\ENDFOR
\RETURN adjacency matrix $A$;
\end{algorithmic}
\end{algorithm}

The \ac{KG} construction algorithm is shown in Algorithm~\ref{alg:Extractgraph}.
$H$ represents the maximum hop count and $N$ represents the number of entities in the final \ac{KG}.
We extract all triples that are related to items involved in the hybrid interaction sequence $S$ within $H$ hops.
We stop extracting more hops if it meets the requirement that there is a path for any given pair of items from both domains.
Finally, we construct the relational adjacency matrix for the extracted \ac{KG}.

Specifically, for the input $S$, we first divide it into two sub-sequences $S_A$ and $S_B$ according to the domain to which the items belongs (see line 1).
We initialize $\vartheta^0_1$ or $\vartheta^0_2$ with the item entities (see line 2).
At each hop, we extract all triples connected to the entities in the current \ac{KG} (i.e., $\vartheta^k_1$ and $\vartheta^k_2$), where $\vartheta^k_1=\{(h,r,t)|(h, r, t)\in \langle E, R, E \rangle$ and $h \in \vartheta^{k-1}_1 \}$, and $\vartheta^k_2=\{(h,r,t)|(h, r, t)\in \langle E, R, E \rangle$ and $h \in \vartheta^{k-1}_2 \}$, $k$ = 1, 2, \ldots, $H$, where $\vartheta^0_1$ = $S_A$ and $\vartheta^0_2$ = $S_B$ (see line 4 to line 5).
If there is a path for any given pair of items from both domains (e.g, $A_{3} \rightarrow e_{8} \rightarrow e_{26} \rightarrow e_{9} \rightarrow B_{6}$) or the hop reaches the maximum hop $H$, we stop extracting other triples from the complete \ac{KG} and construct the relational adjacency matrix (see line 6 to line 12).
Otherwise, we continue to extract other related triples.
When constructing the adjacency matrix, we limit the number of entities in the \ac{KG} to $N$.
Therefore, we need to select some triples if the number of entities in all related triples is larger than $N$ (see line 8).
To do so, we first gather all the entities that connect a pair of item entities from two domains, e.g., $e_{8}, e_{26}, e_{9}$ in the path $A_{3} \rightarrow e_{8} \rightarrow e_{26} \rightarrow e_{9} \rightarrow B_{6}$.
Then, we select the entities according to their smallest distance w.r.t.\ any item from the two domains until the number of all entities meets $N$.
After that, we construct the relational adjacency matrix $A$ based on the selected triples (see line 10).

\subsection{\ac{MIFN}}
\label{section:walkthrough}
In the following subsections, we will demonstrate the details of \ac{MIFN}.
Generally, \ac{MIFN} models $P(A_{i+1}|S, \langle E, R \rangle)$ and $P(B_{j+1}|S, \langle E, R \rangle)$ (see Eq.~\ref{general_equations}) by taking two recommendation modes into consideration, as shown as in Eq.~\ref{model}:
\begin{equation}
\label{model}
\begin{split}
    P&(A_{i+1}|S,\langle E, R \rangle) = \\
     &P(M\_S_{A}|S,\langle E, R \rangle)P(A_{i+1}|M\_S_{A},S,\langle E, R \rangle)+
    P(M\_G_{A}|S,\langle E, R \rangle)P(A_{i+1}|M\_G_{A},S,\langle E, R \rangle)  \\
    P&(B_{j+1}|S,\langle E, R \rangle)  = \\
    & P(M\_S_{B}|S,\langle E, R \rangle)P(B_{j+1}|M\_S_{B},S,\langle E, R \rangle)+
    P(M\_G_{B}|S,\langle E, R \rangle)P(B_{j+1}|M\_G_{B},S,\langle E, R \rangle). \\ 
\end{split}
\end{equation}
Here, $M\_S$ and $M\_G$ denote \textit{sequence mode} and \textit{graph mode}, which make recommendations at the user behavior level and the knowledge level, respectively.
$P(M\_S_{A}|S,\langle E, R \rangle)$ and $P(M\_G_{A}|S,\langle E, R \rangle)$ represent the probabilities under the \textit{sequence mode} and the \textit{graph mode} in domain $A$, respectively, $P(A_{i+1}|M\_S_{A},S,\langle E, R \rangle)$ and $P(A_{i+1}|M\_G_{A},S,\langle E, R \rangle)$ refer to the probabilities of recommending the next item $A_{i+1}$ under the \textit{sequence mode} and \textit{graph mode} given a hybrid interaction sequence $S$ and the \ac{KG} triples $\langle E, R \rangle$.
The same definitions apply to domain $B$.

\begin{figure*}[ht]
\centering
\includegraphics[clip,trim=0mm 0mm 45mm 0mm,width=\textwidth]{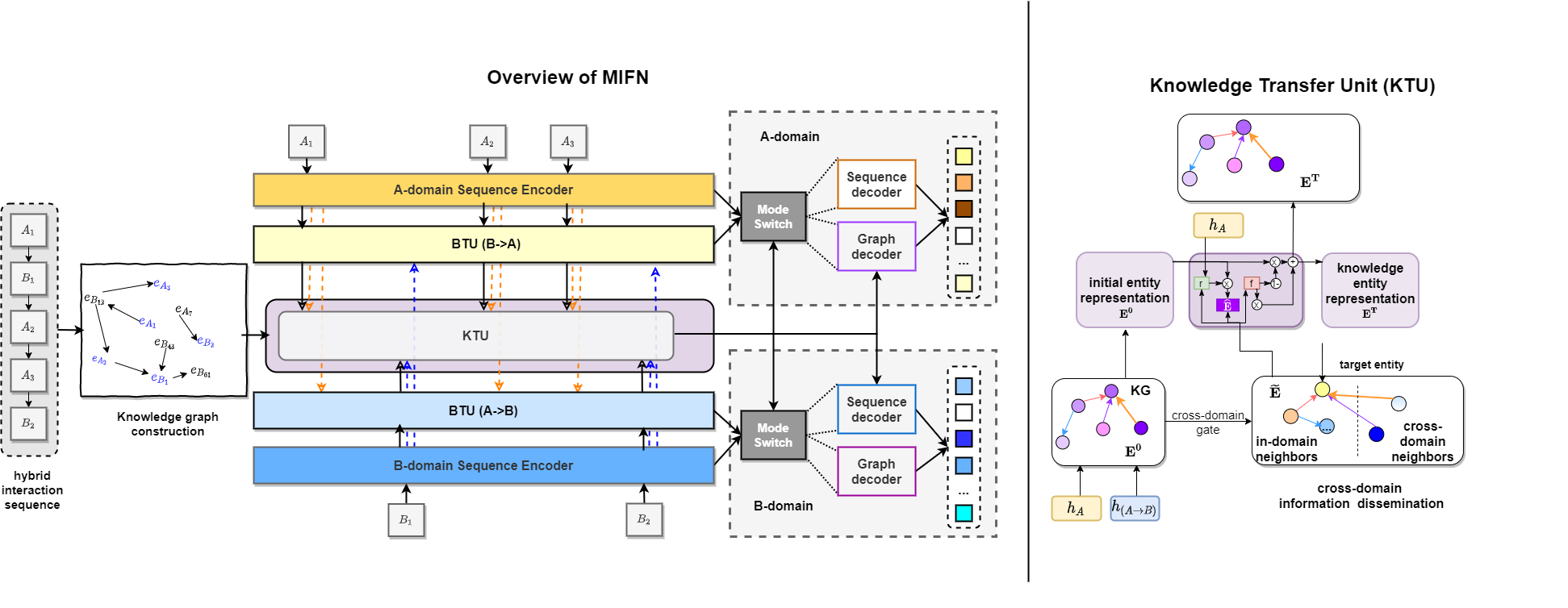}
\caption{An overview of \ac{MIFN}.
Section~\ref{section:walkthrough} contains a walkthrough of the model.
}
\label{03-1}
\end{figure*}

\noindent%
As shown in the left side of Figure~\ref{03-1}, \ac{MIFN} consists of four main components: a sequence encoder, a \acf{BTU}, a \acf{KTU}, and a mixed recommendation decoder. 
The sequence encoder encodes the interacted item sequence into a sequence of item representations.
The \ac{BTU} takes the representations from the source domain as input, extracts behavioral information flow, and transfers it to the target domain.
The \ac{KTU} aims to grasp useful knowledge from the \ac{KG} and propagates it to both domains.
The mixed recommendation decoder contains two decoders w.r.t. \textit{graph mode} and \textit{sequence mode}, respectively.
The \textit{graph recommendation decoder} evaluates the probability for all candidate items from the \ac{KG}, corresponding to Eq.~\ref{g_rec}.
The \textit{sequence recommendation decoder} evaluates the probability of clicking items, corresponding to Eq.~\ref{s_rec}.

\subsection{Sequence encoder}
\label{sequence_encoder}
As with existing studies \cite{GRU4rec, hidasi2016parallel}, we use an \ac{RNN} to encode the sub-sequences $S_A$ and $S_B$. 
Here, we employ a \ac{GRU} as the recurrent unit.
The initial state of the \acp{GRU} is set to zero vectors, i.e., $h_0=0$.
After that, we can obtain $H_A=\{h_{A_1}$, $h_{A_2}$, \ldots, $h_{A_i}$, \ldots, $h_{A_n}\}$ for domain $A$, and $H_B=\{h_{B_1},h_{B_2},\ldots,h_{B_j},\ldots,h_{B_m}\}$ for domain $B$.
Each $h_{A_i}$ or $h_{B_j}$ is the item representation of an item $A_i$ in sequence $S_A$ or $B_j$ in $S_B$.

\subsection{Behavior transfer unit}
\label{behavior_transfer_unit}

The outputs $H_A$ and $H_B$ from the sequence encoder are representations of user behavior in single domains.
It has been shown that there is connection between $H_A$ and $H_B$~\cite{ma2019pi}.
For example, a user who has read the \textit{Harry Potter} book (e.g., ``\textit{Harry Potter and the Philosopher's Stone}'' or ``\textit{Harry Potter and the Chamber of Secrets}'' and so on) has also watched the ``\textit{Pirates of the Caribbean}'' movie within the same time period.
Based on behavioral information from both domains, it is easier for the model to infer that the user might like some magic and fantastic movies and books.

To achieve this, we employ the \ac{BTU} to model the flow of behavioral information from domain $A$ to domain $B$, i.e., $h_{A_{i}\rightarrow B}$, as follows:
\begin{equation}
\begin{split}
\label{distill gate}
f_{A_i} &{} = \sigma(W_{f_A}\cdot h_{A_i}+W_{f_B}\cdot h_{B_j}+ W_{f}\cdot h_{A_{i-1}\rightarrow B}+ b_{f}) \\
\widehat{h_{A_i}} & {} = \tanh({W_h \cdot h_{A_i}+ U_h\cdot h_{A_{i-1}\rightarrow B} + b_h }) \\
h_{A_{i}\rightarrow B} & {} = f_{A_i} \odot \widehat{h_{A_i}} + (1-f_{A_i})\odot h_{A_{i-1}\rightarrow B},
\end{split}
\end{equation}
where $h_{A_i}$ and $h_{B_j}$ are the representations of domain $A$ and $B$ at timestamp $i$ and $j$, respectively.
$f_{A_i}$ measures the degree of connection between these two representations $h_{A_i}$ and $h_{B_j}$ from both domains, which employs the gate mechanism to control how much information is to be transferred from domain $A$ to domain $B$.
$\widehat{h_{A_i}}$ is the updated representation of the current input.
$W_A$ and $W_B$ are the parameters, $b_f$ is the bias term, $\odot$ indicates element-wise multiplication.
$h_{A_{i}\rightarrow B}$ can be seen as a combination of $\widehat{h_{A_i}}$ and $h_{A_{i-1}\rightarrow B}$ balanced by $f_{A_i}$.
Note that the \ac{BTU} can be applied bidirectionally from ``domain $A$ to domain $B$'' and ``domain $B$ to domain $A$''.
Here, we take the ``domain $A$ to domain $B$'' direction and achieve recommendations in domain $B$ as an example.

$h_{A_{i}\rightarrow B}$ is the information extracted from domain $A$, which is ready to be transferred to domain $B$.
Since it still belongs to domain $A$, we employ an \ac{RNN} structure to transfer it to domain $B$ as follows:
\begin{equation}
\begin{split}
\label{behavior transfer}
&h_{(A \rightarrow B)_i} = GRU(h_{(A\rightarrow B)_{i-1}}, h_{A_i\rightarrow B}).\\
\end{split}
\end{equation}
After that, we can obtain the transferred behavioral representation $h_{(A \rightarrow B)_i}$ in domain $B$ at time step $i$.

\subsection{Knowledge transfer unit}
\label{gt_sec}
The \ac{BTU} only models the flow of behavioral information.
We hypothesize that this may be not enough for the model to be able to encode the connection between items from the two domains in some cases.
For example, if there is no knowledge indicating that both ``\textit{Pirates of the Caribbean}'' and ``\textit{Harry Potter}'' belong to fantasy, it is difficult for the model to capture the connection between them solely based on behavioral information.
To better transfer the information of items from both domains, we propose the \ac{KTU}, as shown in the right side of the Figure~\ref{03-1}.

For each hybrid interaction sequence $S$, we get the item representations $\{h_{A_1}$,  $h_{A_2}$, \ldots, $h_{A_i}, \ldots\}$ from the sequence encoder (Section~\ref{sequence_encoder}) and the transferred behavioral representations $\{h_{(A \rightarrow B)_1}$, $h_{(A \rightarrow B)_2}, \ldots, h_{(A \rightarrow B)_i}, \ldots\}$ from the \ac{BTU} (Section~\ref{behavior_transfer_unit}).
We use the item representation of the last time step to denote the sequence representation $h_{A}$ and the transferred behavioral representation $h_{(A \rightarrow B)}$.
We also obtain the relational adjacency matrix $A$ of the \ac{KG}, which consists of $N$ entities and the corresponding relations (\S\ref{graph_constructor}).
Here, we represent these $N$ entities as $E$, the corresponding relations are represented as $R$.

We initialize all entities in the \ac{KG} and we can get the initialized entity representations $h_{E^0}$.
That is, for each entity $e_k \in E$, the initialized entity representation is $h_{e_k^0} \in h_{E^0}$.
Then we learn a transferred entity representation $h_{e_k^T} \in h_{E^T}$ for each entity $e_k$ by leveraging the relations in the \ac{KG}, as shown in Eq.~\ref{gated_transfer}:
\begin{equation}
\begin{split}
\label{gated_transfer}
r &= \sigma(W_r \cdot h_{e_k^0} + U_r \cdot \widetilde{h_{e^L_k}} + b_r) \\
f &= \sigma(W_f \cdot h_{e_k^0} + U_f \cdot \widetilde{h_{e^L_k}} + V_f \cdot h_{A}+ b_f) \\
\widehat{h_{e_k}} & = \tanh(W_h \cdot \widetilde{h_{e^L_k}} + U_h\odot (r \odot h_{e_k^0})) \\
h_{e_k^T} & = (1-f) \odot h_{e_k^0} + f \odot \widehat{h_{e_k}}. \\
\end{split}
\end{equation}
The explanations for the main parts of Eq.~\ref{gated_transfer} are as follows:
\begin{enumerate}[label=(\roman*)]
\item \textbf{Gated functions}. $r$ and $f$ are the update gate function and the forget gate function, which aim to regulate how much of the update information should be propagated.
$W_r$, $U_r$, $W_f$, $U_f$ and $V_f$ are the parameters; $b_r$ and $b_f$ are the bias term.

\item \textbf{Candidate knowledge transfer representation}. 
$\widehat{h_{e_k}}$ is the candidate knowledge transfer representation, which is calculated based on the cross-domain disseminated entity representation $ \widetilde{h_{e^L_k}}$ at the $L$-th hop layer (we will explain this later in Eq.~\ref{gt multi-hop reasoning}) and the updated entity representation $r \odot h_{e_k^0}$.
$W_h$ and $U_h$ are the parameters, $\odot$ indicates element-wise multiplication.

\item \textbf{Transferred entity representation}.
The transferred entity representation $h_{e_k^T}$ is a combination of the initialized entity representation $h_{e_k^0}$ and the candidate knowledge transfer representation $\widehat{h_{e_k}}$ balanced by the forget gate $f$, where the information among entities has been disseminated through $L$ hop layers.
\end{enumerate}

\noindent%
Graph convolutional techniques are commonly used to disseminate information among entities based on their relations~\cite{kipf2016semi,KGCN,RippleNet}.
However, in the cross-domain scenario that we consider, the information disseminated by entities from different domains is different.
Hence, we propose a \acl{CDGCM} that can distinguish entities from different domains and adopt different modeling methods to disseminate their information so as to get better entity representation.
In this manner, the information in the \ac{KG} is disseminated between both domains, which can be considered as a flow of knowledge in the hybrid interaction sequence.
Suppose the information can be disseminated within $L$ hop layers.
At the $l$-th hop layer ($0 \leq l \leq L$), the information of each entity and its cross-domain neighbor entities via various relations will be disseminated to the next hop layer $l+1$ and is used to update the entity representation. 
The process of cross-domain information dissemination is defined in Eq.~\ref{gt multi-hop reasoning}:
\begin{equation}
\begin{split}
\label{gt multi-hop reasoning}
\mbox{}\hspace*{-.2cm}
\widetilde{h_{e^{l+1}_k}} = \sigma \left[ f_0(\widetilde{h_{e^{l}_k}}) 
+ \frac{1}{|N_i(k)|} \sum_{r\in R} \sum_{p\in N_i(k)} f_i(\alpha_p \cdot \widetilde{h_{e^l_{p}}})
+ \frac{1}{|N_c(k)|}\sum_{r\in R} \sum_{q\in N_c(k)} f_c(\beta_q \cdot \widetilde{h_{e^l_{q}}} + c_q \cdot \widetilde{h_{e^l_{q}}})
\right],
\hspace*{-.1cm}\mbox{}
\end{split}
\end{equation}
where $e_k$ is any entity in the entity set $E$; $N_i(k)$ is the neighbor entity set of entity $e_k$ from the same domain as $e_k$;
$N_c(k)$ is the neighbor entity set of entity $e_k$ from the complementary domain.
$f_0$, $f_i$ and $f_c$ represent the transformation functions for initialization, in-domain and cross-domain respectively;
$\alpha_p$ is the in-domain attention weight calculated between each entity $e_p \in N_i(k)$ and the sequence representation $h_{A}$;
$\beta_q$ is the cross-domain attention weight calculated between each entity $e_q \in N_c(k)$ and the transferred behavioral representation $h_{(A \rightarrow B)}$;
$c_q$ shows the sum of similarities between entity $e_q$ and each entity $e_p \in N_i(k)$, which is defined as: $c_q = \sum_{p\in N_i(k)} e_p \cdot e_q$;
$\widetilde{h_{e^l_k}}$ denotes the cross-domain disseminated representation of entity $e_k$ at the $l$-th hop layer, which aggregates the information from itself and cross-domain neighbor entities as the new representation for the next hop layer.
At the first hop layer, the cross-domain disseminated representation is assigned by the gated entity representation, i.e., $\widetilde{h_{e^0_k}} = h_{e_{A_k}}$ if the entity $e_k$ is from domain $A$, otherwise $\widetilde{h_{e^0_k}} = h_{e_{B_k}}$: 
%
\begin{equation}
\begin{split}
\label{cross-gate}
c & = \sigma\left(W_c \cdot concat[h_{A},h_{(A \rightarrow B)},h_{E_0}]  + b_c\right) \\
h_{e_{A_k}} & = c \odot \left(\alpha_{A_k} \cdot h_{e_{A_k}^0}\right) \\
h_{e_{B_k}} & = (1-c) \odot \left(\beta_{B_k} \cdot h_{e_{B_k}^0}\right), \\
\end{split}
\end{equation}
where $W_c$ is the parameter; $b_c$ is the bias term; $h_{e_{A_k}}$ is the gated entity representation of entity $e_{A_k} \in E_A$, $h_{e_{B_k}}$ is the gated representation of entity $e_{B_k} \in E_B$.
$c$ is a cross-domain information gate to handle the situation where the proportion of entities from different domains in the \ac{KG} is different (e.g., when there are 1000 entities from domain $A$, but only 10 entities from domain $B$).
So we define the gated entity representations $h_{e_{A_k}}$ for domain $A$ and $h_{e_{B_k}}$ for domain $B$ based on their initial entity representations respectively, which aims to balance information from both domains.
$\alpha_{A_k}$ is the attention weight of $e_{A_k}$ for sequence representation $h_{A}$; $\beta_{B_k}$ is the attention weight of $e_{B_k}$ for the transferred behavioral representation $h_{(A \rightarrow B)}$.
These attention weights act as an information controller to identify entities of different importance in the \ac{KG}, the definitions of which are shown in Eq.~\ref{softmask}:
\begin{equation}
\begin{split}
\label{softmask}
s_{A_k} & = \mathbf{v_1}^\mathsf{T}\tanh{\left(W_{A_1}\cdot h_{A}+W_{A_2}\cdot h_{e_{A_k}^0} \right)} \\
\alpha_{A_k} & = \frac{\exp{(s_{A_k})}}{\sum_{i}\exp{(s_{A_i}})} \\
s_{B_k} & = \mathbf{v_2}^\mathsf{T}\tanh{\left(W_{B_1}\cdot h_{(A \rightarrow B)} +W_{B_2}\cdot h_{e_{B_k}^0} \right)} \\
\beta_{B_k} & = \frac{\exp{(s_{B_k})}}{\sum_{j}\exp{(s_{B_j}})}, \\
\end{split}
\end{equation}
where $h_{e_{A_k}^0}$ and $h_{e_{B_k}^0}$ are the initialized entity representations of entity $e_{A_k}$ and $e_{B_k}$ respectively.
$h_A$ is the sequence representation and $h_{(A \rightarrow B)}$ is the transferred behavioral representation as mention above. 
$\mathbf{v_1}$, $\mathbf{v_2}$, $W_{A_1}$, $W_{A_2}$, $W_{B_1}$ and $W_{B_2}$ are learnable parameters.

\subsection{Mode switch}
\label{mode_switch_sec}
Recall that $P(M\_S_{B}|S,\langle E, R \rangle)$ and $P(M\_G_{B}|S,\langle E, R \rangle)$ are the probabilities of conducting recommendations under \textit{sequence mode} and \textit{graph mode}, respectively.
We model the mode switch as a binary classifier.
Specifically, we first combine the sequence representation $h_{B}$, the transferred behavioral representation $h_{(A\rightarrow B)}$ and the sum of all transferred entity representation $\sum_{k}(h_{e_k^T})$ (where $h_{e_k^T} \in h_{E^T}$).
Then, we employ a softmax regression to transform the total representation into the mode probability distributions, as follows:
\begin{equation}
\begin{split}
\label{gen_mode}
P(M\_G_B|S, \langle E, R \rangle),P(M\_S_B|S, \langle E, R \rangle) = \text{softmax}\left(W_m\cdot concat\left[h_{B}, h_{(A\rightarrow B)}, \sum_{k}(h_{e_k^T})\right]+b_m\right),
\end{split}
\end{equation}
where $W_m$ is the weight matrix and $b_m$ is the bias term.

\subsection{Graph recommendation decoder}
The graph recommendation decoder evaluates the probabilities of recommending items involved in the \ac{KG}.
Here we directly use the representations of entities to learn the attention weights and take these weights as the final predicted recommendation probability.
The recommendation probability for each item $B_{j+1} \in E_B$ is computed as follows:
\begin{equation}
\begin{split}
\label{g_rec}
&P(B_{j+1}| M\_G_{B},S,\langle E, R \rangle)=
\begin{cases}
0 & \text{if }{B_{j+1} \notin \mathbbm{B}} \\
\frac{\exp(h_{e_{B_{j+1}}^T})}{\sum_{k}^{m} \exp(h_{e_k^T})} & \text{if }{B_{j+1} \in \mathbbm{B}},
\end{cases}
\end{split}
\end{equation}
where $h_{e_{B_{j+1}}^T}$ is the transferred entity representation of $e_{B_{j+1}}$ (corresponding to $h_{e_k^T}$ in Eq.~\ref{gated_transfer} when $e_k$ is an item entity from domain $B$).
$m$ is the number of the item set $\mathbbm{B}$.
Note that $e_{B_{j+1}} \in E_B$ is an item entity corresponding to item $B_{j+1} \in \mathbbm{B}$.
The recommendation probabilities are set to zero for those items that do not exist in $\mathbbm{B}$.

\subsection{Sequence recommendation decoder}
The sequence recommendation decoder evaluates the probabilities of items in the sequence item set.
We first concatenate the sequence representation $h_{B}$ and the transferred behavioral representation $h_{(A\rightarrow B)}$ into the hybrid representation $c_S$, i.e., $c_S = \begin{bmatrix} h_{B}, h_{(A\rightarrow B)}\\ \end{bmatrix}^\mathsf{T}$.
Then, the recommendation probability for each item $B_{j+1} \in \mathbbm{B}$ is computed as follows:
\begin{equation}
\begin{split}
\label{s_rec}
&P(B_{j+1}|M\_S_B,S,\langle E, R \rangle) =
\begin{cases}
0 & \text{if }{B_{j+1} \notin \mathbbm{B}} \\
\frac{\exp(W_{j+1} c_S + b_I)}{\sum_{k}^{m} \exp(W_k c_S + b_I)} & \text{if }{B_{j+1} \in \mathbbm{B}},
\end{cases}
\end{split}
\end{equation}
where $W_k \in W_I$ is the weight matrix, $b_I$ is the bias term.
The recommendation probabilities are set to zero for those items that do not exist in the item set $\mathbbm{B}$.

\subsection{Objective function}
Our goal is to maximize the prediction probability for each domain given a hybrid interaction sequence.
Therefore, we define the negative log-likelihood loss function as follows:
\begin{align}
L_R(\theta)= L_{R_A}(\theta)+L_{R_B}(\theta),
\label{total loss}
\end{align}
where $\theta$ are all parameters in \ac{MIFN}.
Specifically, $L_{R_A}(\theta)$ and $L_{R_B}(\theta)$ can be derived as follows:
\begin{equation}
\begin{split}
L_{R_A}(\theta) & = -\frac{1}{|\mathbb{S}|} \sum_{S \in \mathbb{S}} \sum_{A_i \in S} \log P(A_{i+1}|S,\langle E, R \rangle) \\
L_{R_B}(\theta) & = -\frac{1}{|\mathbb{S}|} \sum_{S \in \mathbb{S}} \sum_{B_j \in S} \log P(B_{j+1}|S,\langle E, R \rangle),
\end{split}
\end{equation}
where $\mathbb{S}$ is the set of all hybrid interaction sequences in our training set, and $P(A_{i+1}|S,\langle E, R \rangle)$ or $P(B_{j+1}|S,\langle E, R \rangle)$ are the next prediction probabilities, which are as defined in Eq.~\ref{model}.

Additionally, \ac{MIFN} incorporates a \textit{mode switch} module to calculate the mode selection probability between \textit{sequence mode} and \textit{graph mode}.
We assume that if an item does not exist in item set, it can just be generated under the \textit{graph mode}.
Here, we can jointly train another mode prediction loss as follows, which adopts the negative log-likelihood loss:
\begin{equation}
\begin{split}
\label{A_mode_loss}
L_{M_A}(\theta) & = -\frac{1}{|\mathbb{S}|} \sum_{S \in \mathbb{S}} \sum_{A_i \in S} \lbrack (1-\mathbbm{1}(A_{i+1} \in \mathbbm{A}))\log P(M\_S_A|S,\langle E, R \rangle) \rbrack \\
L_{M_B}(\theta) & = -\frac{1}{|\mathbb{S}|} \sum_{S \in \mathbb{S}} \sum_{B_j \in S} \lbrack (1-\mathbbm{1}(B_{j+1} \in \mathbbm{B}))\log P(M\_S_B|S,\langle E, R \rangle) \rbrack \\
L_M(\theta) & = L_{M_A}(\theta)+L_{M_B}(\theta), \\
\end{split}
\end{equation}
where $\mathbbm{1}(item \in item set)$ is the indicator function that equals 1 if this item is in the item set and 0 otherwise.
$L_M(\theta)$ is the total mode loss for domain $A$ and $B$.

Finally, we adopt a joint-learning strategy, and the final loss combines both recommendation loss and mode loss:
\begin{equation}
\begin{split}
\label{A_mode_loss}
&L(\theta) = L_{R}(\theta)+L_{M}(\theta). \\
\end{split}
\end{equation}


\section{Experimental Setup}

\subsection{Research questions}
We evaluate \ac{MIFN} on four e-commerce datasets.
We aim to answer the following questions in our experiments:
\begin{enumerate}[leftmargin=*, nosep, label=(RQ\arabic*)]
\item How does \ac{MIFN} perform compared with the state-of-the-art methods in terms of Recall and MRR? (See Section~\ref{result_sec}.)
\item Does the \ac{KTU} help to improve the performance of recommendations? And does the performance differ from the situation when we only allow for the flow of behavioral  information? (See Section~\ref{subsection:ablation}.)
\item Does the knowledge graph construction method have a big effect on the overall recommendation results? (See Section~\ref{algorithm_influence}.)
\item Is \ac{MIFN} able to provide better recommendations by incorporating the flow of knowledge across domains? (See Section~\ref{case_study}.) 
\end{enumerate}

\subsection{Datasets}
We conduct experiments on the Amazon e-commerce collection,\footnote{\url{https://www.amazon.com/}} which consists of user interactions (e.g., userid, itemid, ratings, timestamps) from multiple domains and some item meta information (e.g., descriptions, images, product associations).
Compared with other recommendation datasets, the Amazon dataset contains overlapping user interactions in multiple domains,  which is suitable for \acf{CDSR}.
Specifically, we pick two pairs of complementary domains ``Movie-Book'' domains and ``Food-Kitchen'' domains for experiments. 
For the ``Movie-Book'' dataset, the ``Movie'' domain contains movie watching records. 
The ``Book'' domain covers book reading records.
For the ``Food-Kitchen'' dataset, the ``Food'' domain contains food purchase records.
The ``Kitchen'' domain contains furniture purchase records.
We follow the settings of \citet{ma2019pi} to process the data.
To satisfy cross-domain characteristics, we first pick users who have interactions in both domains.
Since we do not target cold-start users or items in this work, we only keep users who have more than 10 interactions and items whose frequency is larger than 10.
To satisfy sequential characteristics which consists of many user interactions within a period of time, we first order the interactions by time for each user, then we split the sequences from each user into several small sequences with each sequence containing interactions within a period, i.e., a month for the ``Movie-Book'' dataset, and a year for the ``Food-Kitchen'' dataset.
We also require that each sequence contains at least 3 items from each domain.
The statistics of the processed datasets are shown in Table~\ref{dataset1}.

\begin{table}[h]
\centering
\caption{Dataset statistics.}
\label{dataset1}
\begin{tabular}{lccccc}
\toprule  
\bf Domain  & \bf \#Items  &\bf \#Train & \bf \#Test & \bf \#Valid & \bf Avg\_Seq\_Len \\
\midrule  
Movie   & 36,845 & \multirow{2}{*}{44,732} &\multirow{2}{*}{19,861} & \multirow{2}{*}{9,274} & \multirow{2}{*}{11.98} \\
Book    & 63,937 & & & &     \\
\midrule  
Food    & 29,207 & \multirow{2}{*}{25,766} &\multirow{2}{*}{17,280,} & \multirow{2}{*}{7,650} & \multirow{2}{*}{\phantom{1}9.91} \\
Kitchen    & 34,886 & & & &     \\
 \bottomrule 
\end{tabular} 
\end{table}

\noindent%
For knowledge graph construction, we use the Amazon product data to mine the relations of the items.\footnote{\url{https://jmcauley.ucsd.edu/data/amazon}}
The data is collected from large-scale user logs, which contain the following relations:
\begin{enumerate*}
\item ``Also\_buy'' (users also buy item $X$ when buying item $Y$.); 
\item ``Also\_view'' (users also view item $X$ when viewing item $Y$.); 
\item ``Buy\_together'' (users buy item $X$ and $Y$ together frequently);
\item ``Buy\_after\_viewing'' (users buy item $X$ after they buy $Y$);
\item ``Is\_the\_same\_category'' (item $X$ and $Y$ belong to the same category).
\end{enumerate*}
Additionally, for the ``Movie-Book'' dataset, we also crawl the relation ``Adapted\_from'' between books and movies from Wikipedia.\footnote{\url{https://en.wikipedia.org/}}
For example, $\langle$movie, Adapted\_from, book$\rangle$ means that the movie is adapted from the book.
We align knowledge entities with Wikipedia titles by fully matching.
The statistics of the knowledge information are shown in Table~\ref{dataset2}.

\begin{table}[h]
\centering
\caption{\ac{KG} statistics.}
\label{dataset2}
\begin{tabular}{lccc}
\toprule 
\bf Domain  & \bf \#Entities & \bf \#Relations  &\bf \#Triples \\
\midrule  
Movie   & \phantom{0}65,418  & \multirow{2}{*}{6} & \phantom{0}3,911,284 \\
Book    & 315,770 &                    & 40,048,795 \\
 \midrule 
Food   & \phantom{0}50,273  & \multirow{2}{*}{5} & \phantom{0}3,822,123 \\
Kitchen  & \phantom{0}82,552 &                    & \phantom{0}7,836,064 \\
 \bottomrule 
\end{tabular} 
\end{table}

\noindent%
For evaluation, we use the last interacted item in each sequence for each domain as the ground truth item, respectively. 
We randomly select 80\% of each user's interactions as the training set, 10\% as the validation set, and the remaining 10\% as the test set.

\subsection{Baseline methods}
We compare the proposed model \ac{MIFN} with baselines from five categories: 
\begin{enumerate*}
\item traditional recommendation methods, 
\item cross-domain recommendation methods, 
\item sequential recommendation methods, 
\item cross-domain sequential recommendation methods, and 
\item knowledge-aware recommendation methods.
\end{enumerate*}

\subsubsection{Traditional recommendation methods.}
We adapt three commonly used traditional recommendation methods to \acp{SR}:
\begin{itemize}[leftmargin=*,nosep]
\item POP: This method recommends the most popular items in which items are ranked based on their popularity. 
It is a simple baseline, but is commonly used owing to its simplicity yet effectiveness~\cite{ncf_he}. 

\item Item-KNN:  This method is inspired by the classical \ac{KNN} model; it looks for items that are similar to other items that have been clicked by a user in the past, where similarity is defined as the cosine similarity between the vector of sequences~\cite{sarwar2001item}. 

\item BPR-MF: This method follows the idea of \ac{MF} with a pairwise ranking objective via stochastic gradient descent~\cite{rendle2009bpr}.
Following~\citet{GRU4rec}, we represent a new sequence by the average latent factors of items that appeared in the sequence so far.
\end{itemize}

\subsubsection{Cross-domain recommendation methods.}
We use two popular cross-domain recommendation methods for comparison.
\begin{itemize}[leftmargin=*,nosep]
\item NCF-MLP++: This is a deep learning based method where the model learns the inner product of the traditional \ac{CF} by using \ac{MLP} in each domain.
The user representations are shared between both domains while item representations are private in each domain, and the final recommendations are aggregated from both domain recommendations probabilities.
We adopt the implementation in \cite{ma2019pi}.

\item Conet: This method transfers information between different domains by a cross-stitch network~\cite{misra2016cross}, where information in each domain is captured by neural \ac{CF} model~\cite{ncf_he}.
\end{itemize}

\subsubsection{Sequential recommendation methods.}
A number of \ac{SR} methods have been proposed in the last few years.
In this work, we construct/select baselines that are fair (use the same information, similar architectures, etc.) to compare with:
\begin{itemize}[leftmargin=*,nosep]
\item GRU4REC: It is the first attempt to use \ac{GRU} for \acp{SR}.
It utilizes session-parallel mini-batch training strategy and employs a ranking-based loss functions~\cite{GRU4rec}. 

\item HRNN: This method combines the extra user's information into \ac{GRU} networks and proposes a hierarchical \ac{RNN} model based on GRU4REC~\cite{HRNN}.

\item NARM: This method takes an attention mechanism into consideration to capture both sequential-level preferences and the user's main purpose~\cite{NARM}.

\item STAMP: This method constructs two network structures to capture a user's general preferences and the current preferences of the last click within the current sequence~\cite{STAMP}.
\end{itemize}

\subsubsection{Cross-domain Sequential recommendation methods.}
\begin{itemize}[leftmargin=*,nosep]
\item $\pi$-Net: This method is the only one that considers cross-domain characteristics for \acp{SR}. We take this as the fairest baseline to compare with. It designs a new gating mechanism to recurrently extract and share useful information across different domains~\cite{ma2019pi}.
\end{itemize}

\subsubsection{Knowledge-aware recommendation methods.}
\begin{itemize}[leftmargin=*,nosep]
\item SRGNN: This method constructs each sequence as a directed graph, where the items in the sequence are entities and transition relationship between adjacent items represents edge which is considered as knowledge information.
By modeling the complex transitions, each session graph and item embeddings of all items involved in each graph can be obtained through gated graph neural networks~\cite{SRGNN}.

\item KSR: This method incorporates \acp{KG} into \acp{SR}, and it combines the sequential user preference captured by an \ac{RNN} and attribute-level preferences captured by Key-Value Memory Networks to get the final representation of user preference~\cite{KSR}.
\end{itemize}

\subsection{Evaluation metrics}
We target the top-K recommendations in this work, so we adopt two widely used ranking-based metrics~\citep{latentfactormodel,apranking,attnetwork_mei,repeatnet}: MRR@K and Recall@K.
Specifically, we report $K$ = 5, 10, 20.
 
\begin{itemize}[leftmargin=*,nosep]
\item Recall: This measures the proportion of the top-K recommended items that are in the evaluation set.
It does not consider the actual rank of the item as long as it is amongst the list of recommend items.
\item MRR: This is the average of reciprocal ranks of the relevant items. 
And the reciprocal rank is set to zero if the ground truth item is not in the list of recommended items. 
MRR takes the order of recommendation ranking into account.
Since each sample has only one ground truth item, we choose MRR as the ranking metric instead of others, e.g., NDCG.
\end{itemize}

\subsection{Implementation details}
For most of the baseline methods, we find the best settings using grid search on the validation set.
For those with too many hyperparameters, we follow the reported optimal hyperparameter settings from the original publications that introduced them.
For our model, the embedding size and the hidden size are set to 256, the hop layer $H$ is set to 2, and the number of entities $N$ in \ac{KG} is set to 200.
We initialize the model parameters randomly using the Xavier method.
We take Adam as our optimization method.
\ac{MIFN} is implemented in TensorFlow and trained on a GeForce GTX TitanX GPU.
The code and dataset used to run the experiments in this paper are available at \url{https://github.com/mamuyang/MIFN}.


\section{Results and Analysis}

\subsection{Overall performance (RQ1)}
\label{result_sec}
We report the results of \ac{MIFN} compared with the baseline methods on the ``Movie-Book'' and ``Food-Kitchen'' datasets.
The results of all methods are shown in Table ~\ref{results_1} and ~\ref{results_2}, respectively.
From the results, we have the following main observations.

\begin{table*}[]
\centering
\caption{Experimental results (\%) on the Amazon (``Movie-Book'') dataset.
\textbf{Bold face} indicates the best result in terms of the corresponding metric.
Significant improvements over the best baseline results are marked with $^\dagger$ (t-test, $p < .05$).
}
\label{results_1}
\small
\begin{tabular}{lcccccccccccc}
\toprule
\multirow{3}{*}{\bf Methods} & \multicolumn{6}{c}{\bf Movie-domain} & \multicolumn{6}{c}{\bf Book-domain}     \\
\cmidrule(r){2-7}\cmidrule{8-13}&
\multicolumn{3}{c}{MRR} & \multicolumn{3}{c}{Recall} & \multicolumn{3}{c}{MRR} & \multicolumn{3}{c}{Recall}\\
\cmidrule(r){2-4}\cmidrule(r){5-7}\cmidrule(r){8-10}\cmidrule{11-13} & @5  & @10 & @20  & @5  & @10  & @20  & @5  & @10  & @20  & @5  & @10  & @20   \\
\midrule
POP &  \phantom{0}0.10    &  \phantom{0}0.11   &  \phantom{0}0.13  &  \phantom{0}0.20  &   \phantom{0}0.29  &  \phantom{0}0.58  &
\phantom{0}0.11 &   \phantom{0}0.14  & \phantom{0}0.16     & \phantom{0}0.20  & \phantom{0}0.44  &  \phantom{0}0.75      \\

BPR-MF  &    \phantom{0}0.51   &  \phantom{0}0.57  &  \phantom{0}0.64  &  \phantom{0}0.89  &  \phantom{0}1.35  & \phantom{0}2.26 & 
\phantom{0}1.44  & \phantom{0}1.64  &  \phantom{0}1.77  & \phantom{0}2.51 & \phantom{0}3.97  &  \phantom{0}5.97     \\

ItemKNN    &   \phantom{0}1.05    & \phantom{0}1.27  &   \phantom{0}1.48  &    \phantom{0}2.11   &  \phantom{0}3.84   &  \phantom{0}6.99    &  \phantom{0}1.35   & \phantom{0}1.64   &  \phantom{0}1.95  &   \phantom{0}2.88   &   \phantom{0}5.10  & \phantom{0}9.69     \\

\midrule
NCF-MLP++ &  \phantom{0}1.64 &  \phantom{0}1.86    &  \phantom{0}2.03    &   \phantom{0}2.95 &  \phantom{0}4.61  &  \phantom{0}7.24 & 
\phantom{0}1.76   &  \phantom{0}1.98 & \phantom{0}2.11   &   \phantom{0}3.20  &   \phantom{0}4.84    &   \phantom{0}7.34  \\

Conet    & \phantom{0}1.43 &  \phantom{0}1.73    &  \phantom{0}2.01    &   \phantom{0}2.83 &  \phantom{0}5.20  &  \phantom{0}9.24 & 
\phantom{0}1.17   &  \phantom{0}1.36 & \phantom{0}1.51   &   \phantom{0}2.13  &   \phantom{0}3.54    &   \phantom{0}5.77  \\
\midrule
GRU4REC   &  12.80 &  12.86 & 12.88 & 13.69 & 14.11  & 14.43  &
13.87 & 13.92   & 13.95 & 14.64 & 15.02 & 15.34  \\
 
HRNN   &  13.38 & 13.43   & 13.45 & 13.95 & 14.25 & 14.58 &
 14.57 & 14.61   & 14.62 & 14.99 & 15.25 & 15.46  \\

NARM  &  13.80  & 13.85  & 13.86  & 14.20 &  14.53    & 14.80 &
15.25   & 15.26  & 15.27  & 15.57  &  15.66 &  15.78    \\

STAMP  &  12.44  & 12.56  & 12.63  & 13.66 & 14.58 & 15.68  &
11.53 &  11.56 & 11.57  & 11.82 & 12.00  &   12.20  \\

\midrule
$\pi$-Net  &  14.49   & 14.52  & 14.54  & 14.88 &  15.10 &  15.37 &
15.75   & 15.76  & 15.77  & 15.94  &  16.02 &  16.09   \\

\midrule
SRGNN   &  11.77   &  11.84  &  11.88 & 12.66   & 13.18 & 13.87  & 
15.12 &  15.14  & 15.15  &  15.46  &  15.61  &  15.77  \\

KSR   & 14.18  & 14.23  & 14.26  & 14.91  & 15.28 & 15.73  &
 15.84 &  15.87  & 15.89 & 16.51 & 16.72 &  16.92  \\

\midrule
\ac{MIFN}-\ac{KTU} &  14.20   &  14.25 &  14.28 & 14.85 & 15.26 & 15.44 & 14.87 & 14.90 & 14.91 & 15.54 & 15.80 & 16.07  \\
\ac{MIFN}+$L_M$ &   14.73  &  14.75 & 14.81  & 14.87    &  15.96   &  16.02   &
 15.75 &  15.97 &  15.99  &  16.87  &  16.96  &  17.05 \\
\textbf{MIFN}  & \textbf{14.84} & \textbf{14.87} &  \textbf{14.88} & \textbf{15.13}  & \textbf{16.34}\rlap{$^\dagger$} & \textbf{16.56}\rlap{$^\dagger$} & 
\textbf{16.05}  & \textbf{16.16} &  \textbf{16.23}\rlap{$^\dagger$} &  \textbf{16.99}\rlap{$^\dagger$}  &  \textbf{17.03}\rlap{$^\dagger$} &  \textbf{17.13}\rlap{$^\dagger$} \\

\bottomrule
\end{tabular}
\end{table*}

First, \ac{MIFN} outperforms the single-domain \ac{SR} methods (e.g., STAMP, NARM) and the knowledge-aware methods (e.g., SRGNN, KSR) on all datasets.
Particularly, on the ``Movie-Book'' dataset, the largest increase over NARM is 7.5\% and 12.4\% in terms of MRR@5 and Recall@10 on the ``Movie'' domain, and on the ``Book'' domain, the largest increase is 6.2\% and 9.1\% in terms of MRR@20 and Recall@5.
And the increase over KSR on the ``Movie'' domain is 4.6\% and 6.9\% in terms of MRR@5 and Recall@10, on the ``Book'' domain, the increase is  2.1\% and 2.9\% in terms of MRR@20 and Recall@5.
On the ``Food-Kitchen'' dataset, the largest increase over NARM is 6.5\% and 9.3\% in terms of MRR@20 and Recall@20 on the ``Food'' domain, and on the ``Kitchen'' domain, the increase is 9.1\% and 12.3\% in terms of MRR@5 and Recall@10.
And the increase over KSR is 2.7\% and 3.9\% in terms of MRR@20 and Recall@10 on the ``Food'' domain, and on the ``Kitchen'' domain, the increase is 1.6\% and 4.1\% in terms of MRR@20 and Recall@20.
These improvements demonstrate that jointly considering both cross-domain behavior and knowledge is helpful for \ac{SR}.

\begin{table*}[]
\centering
\caption{Experimental results (\%) on the Amazon (``Food-Kitchen'') dataset.
\textbf{Bold face} indicates the best result in terms of the corresponding metric.
Significant improvements over the best baseline results are marked with $^\dagger$ (t-test, $p < .05$). 
}
\label{results_2}
\small
\begin{tabular}{lcccccccccccc}
\toprule
\multirow{3}{*}{\bf Methods} & \multicolumn{6}{c}{\bf Food-domain recommendation} & \multicolumn{6}{c}{\bf Kitchen-domain recommendation}     \\
\cmidrule(r){2-7}\cmidrule{8-13}&
\multicolumn{3}{c}{MRR} & \multicolumn{3}{c}{Recall} & \multicolumn{3}{c}{MRR} & \multicolumn{3}{c}{Recall}\\
\cmidrule(r){2-4}\cmidrule(r){5-7}\cmidrule(r){8-10}\cmidrule{11-13} & @5  & @10 & @20  & @5  & @10  & @20  & @5  & @10  & @20  & @5  & @10  & @20   \\
\midrule
POP &  \phantom{0}0.40    &  \phantom{0}0.49   &  \phantom{0}0.56  &  \phantom{0}0.82  &   \phantom{0}1.50  &  \phantom{0}2.15  &
\phantom{0}0.22 &   \phantom{0}0.25  & \phantom{0}0.27     & \phantom{0}0.40  & \phantom{0}0.60  &  \phantom{0}1.47      \\

BPR-MF  &    \phantom{0}0.82   &  \phantom{0}0.87  &  \phantom{0}0.94  &  \phantom{0}1.35  &  \phantom{0}1.82  & \phantom{0}2.61 & 
\phantom{0}0.41  & \phantom{0}0.47 &  \phantom{0}0.50  & \phantom{0}0.62 & \phantom{0}1.04  & \phantom{0}1.47     \\

ItemKNN    &   \phantom{0}1.55    & \phantom{0}1.98  &   \phantom{0}2.43  &    \phantom{0}3.28   &  \phantom{0}6.70   &  12.47    &
\phantom{0}1.13   & \phantom{0}1.44   &  \phantom{0}1.90  &   \phantom{0}2.60   &   \phantom{0}4.77  & 11.08    \\

\midrule
NCF-MLP++ &  \phantom{0}2.01  & \phantom{0}2.24 & \phantom{0}2.42 & \phantom{0}3.45 & \phantom{0}5.19 & \phantom{0}8.02 &
\phantom{0}0.87 & \phantom{0}1.03  & \phantom{0}1.17 & \phantom{0}1.72 & \phantom{0}3.00 & \phantom{0}4.99  \\
 
Conet    &  \phantom{0}3.38  & \phantom{0}3.64  & \phantom{0}3.82 &  \phantom{0}5.07 & \phantom{0}7.07 & \phantom{0}9.73  &
 \phantom{0}3.30 &  \phantom{0}3.55  & \phantom{0}3.71 & \phantom{0}5.09 & \phantom{0}7.03 &  \phantom{0}9.47  \\
\midrule
GRU4REC   &  \phantom{0}8.10  &  \phantom{0}8.23   & \phantom{0}8.29 &  \phantom{0}9.46 & 10.38  &  11.26  &
\phantom{0}8.36 & \phantom{0}8.39  & \phantom{0}8.41 & \phantom{0}8.70 & \phantom{0}8.93 &  \phantom{0}9.25  \\
 
HRNN   &   \phantom{0}7.22  &  \phantom{0}7.35  &\phantom{0}7.45   &  \phantom{0}8.49 & \phantom{0}9.47   &  10.93  &
 \phantom{0}7.81 &  \phantom{0}7.86   & \phantom{0}7.88 & \phantom{0}8.29 & \phantom{0}8.61 &  \phantom{0}9.12  \\

NARM  &  \phantom{0}9.43  &  \phantom{0}9.54  &\phantom{0}9.62   &  10.34 & 11.86   &  12.23  &
 \phantom{0}8.41 &  \phantom{0}8.44   & \phantom{0}8.46 & \phantom{0}8.69 & \phantom{0}8.91 &  \phantom{0}9.21    \\

STAMP  &  \phantom{0}9.28  &  \phantom{0}9.38  &\phantom{0}9.44   &  10.22 & 10.91   &  11.81  &
 \phantom{0}8.52 &  \phantom{0}8.55   & \phantom{0}8.57 & \phantom{0}8.81 & \phantom{0}9.05 &  \phantom{0}9.28    \\

\midrule
$\pi$-Net  &  \phantom{0}9.56  &  \phantom{0}9.67  &\phantom{0}9.75   &  10.59 & 10.46   &  12.54  &
 \phantom{0}8.57 &  \phantom{0}8.60   & \phantom{0}8.62 & \phantom{0}8.89 & \phantom{0}9.12 &  \phantom{0}9.42  \\

\midrule
SRGNN   &  \phantom{0}7.31  &  \phantom{0}7.49  &\phantom{0}7.60   &  \phantom{0}8.68 &  10.02   &  11.67  &
 \phantom{0}7.20 &  \phantom{0}7.26   & \phantom{0}7.29 & \phantom{0}7.90 & \phantom{0}8.36 &  \phantom{0}8.87   \\

KSR   &  \phantom{0}9.79 &  \phantom{0}9.91  &\phantom{0}9.98   &  10.82 & 11.78   &  12.77  &
 \phantom{0}9.03 &  \phantom{0}9.07   & \phantom{0}9.08 & \phantom{0}9.39 & \phantom{0}9.62 &  \phantom{0}9.92  \\
\midrule
\ac{MIFN}-\ac{KTU} &  \phantom{0}9.43   &  \phantom{0}9.65 & \phantom{0}9.83  & 10.50 & 11.16 & 12.53 &
 \phantom{0}8.29 &  \phantom{0}8.33  & \phantom{0}8.36 & \phantom{0}8.95 & \phantom{0}9.17 & \phantom{0}9.52  \\
\ac{MIFN}+$L_M$ &  \phantom{0}9.86   & \phantom{0}9.88 & 10.03  & 10.93  &  11.96   &  13.14   &
\phantom{0}9.05  &  \phantom{0}9.17 & \phantom{0}9.18 & \phantom{0}9.28 & \phantom{0}9.89 & 10.24 \\
\textbf{MIFN}  &  \textbf{\phantom{0}9.91}   &  \textbf{10.16}  & \textbf{10.25}  & \textbf{11.20}  & \textbf{12.25}  & \textbf{13.27}\rlap{$^\dagger$}  &
 \textbf{\phantom{0}9.18} &  \textbf{\phantom{0}9.21}   &  \textbf{\phantom{0}9.23} & \textbf{\phantom{0}9.72}\rlap{$^\dagger$} & \textbf{10.01}\rlap{$^\dagger$} & \textbf{10.33}\rlap{$^\dagger$} \\

\bottomrule
\end{tabular}
\end{table*}

Second, \ac{MIFN} outperforms the \acl{CDSR} baseline $\pi$-Net, which just makes use of information at user behavior level.
Specifically, \ac{MIFN} outperforms $\pi$-Net in terms of all metrics on both domains.
It demonstrates that considering both knowledge \emph{and} user behavior level information is better than only behavioral information.
Meanwhile, it also proves the effectiveness of the \ac{KTU} module of \ac{MIFN}.
With this module, \ac{MIFN} is able to capture cross-domain knowledge and conduct information transfer in the \ac{KG} so as to learn better sequence representations.

Third, we can observe that the results of \ac{MIFN} in the ``Book'' domain are better than those in the ``Movie'' domain on the ``Movie-Book'' dataset.
We believe that this is because the data is less sparse in the ``Book'' domain compared to that in the ``Movie'' domain.
And the results in the ``Food'' domain are better than the ``Kitchen'' domain on the ``Food-Kitchen'' dataset.
Again, we think it is because the data sparsity difference as the users have more interactions in the ``Food'' domain than in the ``Kitchen'' domain.
With more interaction data, the models can identify more user preference characteristics in the dense domain so as to transfer it to the sparse domain through both the user behavioral information flow and the knowledge information flow.

Fourth, $\pi$-Net outperforms other sequential baselines, which means that cross domain information is beneficial to both domains.
At the same time, knowledge aware methods outperform other sequential baselines, which also means that knowledge information can improve recommendation performance.
Furthermore, it seems that considering knowledge is more useful than modeling cross-domain characteristics, as KSR slightly outperforms $\pi$-Net.

Fifth, the sequential methods achieve much better results than traditional methods and cross-domain methods.
This is because \ac{RNN}-based methods are able to capture the sequential characteristics and can obtain the better representations while the traditional methods neglect this information.
Besides, it seems that STAMP obtains lower results than the sequential method NARM in the ``Movie-Book'' dataset, while it performs better in the ``Kitchen'' domain of ``Food-Kitchen'' dataset.
We believe that this is because of differences in the datasets, e.g., we found that the user preferences in the kitchen domain are relatively more focused.
And the knowledge aware method SRGNN performs worse than most sequential methods.
This is because that SRGNN just employs the transition relation between adjacent items to construct the graph so as to get the representations of different items, however it does not consider the relation between the non-adjacent items (the other sequential methods do consider this), which may also affect item representations.

\subsection{Ablation study (RQ2)}
\label{subsection:ablation}
To verify the effectiveness of the proposed modules, we design the ablation study to compare several model variants.
The results are listed in Table~\ref{results_1} and Table~\ref{results_2}.
\begin{enumerate}[label=(\arabic*)]
\item \ac{MIFN} is the best performing variant, which includes both the \ac{BTU} and \ac{KTU} modules, and trained with the recommendation loss $L_R$ only.
\item \ac{MIFN}-\ac{KTU} is \ac{MIFN} without the \ac{KTU} module and performs information transfer only at the level of behavioral information; 
\item \ac{MIFN}+$L_M$ is \ac{MIFN} by adding the mode switch loss.
\end{enumerate}
First, by removing \ac{KTU}, the performance of \ac{MIFN}-\ac{KTU} is dramatically less than that of \ac{MIFN}, which confirms that considering knowledge flow can improve the cross-domain recommendations.
In addition, the results of \ac{MIFN}-\ac{KTU} are worse than those of the knowledge-aware method KSR, while \ac{MIFN} outperforms KSR on all domains.
This indicates that the \ac{KTU} module is able to make good use of the cross-domain knowledge and is able to better capture user preferences by modeling the cross-domain knowledge flow.

Second, if we jointly train the recommendation loss $L_R$ and the mode switch loss $L_M$, the performance drops a little but its performance is still than that of the baselines.
The switch loss $L_M$ assumes that if the next item does not exist in the item set, it must be recommended under the \textit{graph mode}, which makes the model tend to recommend items existing in the graph.
However, since there already exists a similar supervision signal in $L_R$, which assumes that each item is recommended under the \textit{graph mode}, the \textit{sequence mode}, or a combination of both.
Further adding the $L_M$ loss introduces unnecessary bias towards \textit{graph mode}.

\subsection{Influence of the knowledge graph construction algorithm (RQ3)}
\label{algorithm_influence}
The number of triples in the complete \ac{KG} is large (see Table~\ref{dataset2}), so we propose a knowledge graph construction method as detailed in Algorithm~\ref{alg:Extractgraph} to build \ac{KG} for each sequence.
To study the effect of the knowledge graph construction method, we design an experiment aimed at analyzing the effect of the ratios of the ground truth items in the constructed \ac{KG} on the final recommendation performance.
We achieve this by simulating and controlling the ratios artificially.
Specifically, we add the ground truth items to the extracted entities according to the specified ratios in advance.
The results are shown in Table~\ref{ana_1} and Table~\ref{ana_2}.

\begin{table*}[]
\centering
\caption{Analysis of the knowledge graph construction algorithm on the Amazon (``Movie-Book'') dataset.
The different ratios represent the different proportions of the predicted ground truth items appeared in the \ac{KG}.}
\label{ana_1}
\begin{tabular}{ccccccccccccc}
\toprule
\multirow{3}{*}{\bf Ratios} & \multicolumn{6}{c}{\bf Movie-domain recommendation} & \multicolumn{6}{c}{\bf Book-domain recommendation}     \\
\cmidrule(r){2-7}\cmidrule{8-13}&
\multicolumn{3}{c}{MRR} & \multicolumn{3}{c}{Recall} & \multicolumn{3}{c}{MRR} & \multicolumn{3}{c}{Recall}\\
\cmidrule(r){2-4}\cmidrule(r){5-7}\cmidrule(r){8-10}\cmidrule{11-13} & @5  & @10 & @20  & @5  & @10  & @20  & @5  & @10  & @20  & @5  & @10  & @20   \\
\midrule
30\%  &   16.46  & 16.83  & 16.99 &  21.27   & 24.33  & 28.97  &
18.51 & 18.70 & 18.85 & 25.82  & 27.32  & 31.38  \\

50\% & 19.53   &  20.43   & 20.83 &  30.33  &   37.07  &  42.67  &
25.04 &   25.68  & 25.98  & 36.08  & 40.75  &  45.12 \\

70\%  &  23.41   &  23.43  &  23.45  &  42.54  &  47.99  & 50.61 & 
27.12  & 28.31 &  28.41  & 48.93 & 56.64  & 57.31    \\

90\%  &  33.45  & 34.99 &  35.09  &  65.81   &   76.83   & 77.97 &
40.01  & 40.86 & 43.00  & 79.83  & 80.50 & 83.86  \\
 
100\%    &  67.04  & 67.56  & 67.69 &  89.57 & 93.30 & 95.15  &
 83.08 &  83.46  & 83.62 & 91.97 & 94.79 &  97.02  \\

\bottomrule
\end{tabular}
\end{table*}

\begin{table*}[]
\centering
\caption{Analysis of the knowledge graph construction algorithm on the Amazon (``Food-Kitchen'') dataset.
The different ratios represent the different proportions of the predicted ground truth items appeared in the \ac{KG}.}
\label{ana_2}
\begin{tabular}{ccccccccccccc}
\toprule
\multirow{3}{*}{\bf Ratios} & \multicolumn{6}{c}{\bf Food-domain recommendation} & \multicolumn{6}{c}{\bf Kitchen-domain recommendation}     \\
\cmidrule(r){2-7}\cmidrule{8-13}&
\multicolumn{3}{c}{MRR} & \multicolumn{3}{c}{Recall} & \multicolumn{3}{c}{MRR} & \multicolumn{3}{c}{Recall}\\
\cmidrule(r){2-4}\cmidrule(r){5-7}\cmidrule(r){8-10}\cmidrule{11-13} & @5  & @10 & @20  & @5  & @10  & @20  & @5  & @10  & @20  & @5  & @10  & @20   \\
\midrule
30\%  &  11.75   &  12.30  & 12.72   &  17.67  &  21.77  &  27.99  &
10.96 &  11.51  & 11.99   & 13.71  & 15.52 &  22.41     \\

50\%  &  22.14   & 22.61   &  22.74  & 35.21   & 38.51   &  42.47  & 
22.75  & 23.20  & 23.74  &  34.56  &  37.86 & 41.64    \\

70\%  &  26.01  & 26.74    &  27.35  & 41.52   & 47.11   & 56.17   &
25.46   & 26.24   &  26.79  & 37.54   &  43.51  & 51.60    \\

90\%  & 36.58  & 37.50 & 38.11 & 59.72 & 66.57 & 75.61 &
31.29 & 32.88  & 33.40 & 55.21 & 66.78 & 74.17  \\
 
100\% & 47.40  & 48.51  &  48.60  & 85.15  & 92.89  & 94.13 &
43.78  & 45.57  & 45.77 &  80.16 & 92.68 & 95.45  \\

\bottomrule
\end{tabular}
\end{table*}

First, we can see that the performance increases as the ratio of ground truth item appeared in the \ac{KG} increases on both datasets.
For instance, when the ratio varies from 30\% to 100\%, the value of MRR@20 increases from 16.99\% to 67.69\% in the ``Movie'' domain and the value of Recall@20 increases from 22.41\% to 95.45\% in the ``Kitchen'' domain.
This demonstrates that a good knowledge graph construction algorithm is of vital importance.

Second, we notice that when the ratio increases linearly, the results do not increase linearly.
On both datasets, the increase is relatively slow from 30\% to 90\%.
However, when we simulate the ratio from 90\% to 100\%, the performance is greatly improved in terms of both MRR and Recall on the ``Movie-Book'' dataset.
We believe that this is because when the ratios reaches a certain value, the model can easily capture the characteristics of recommended items from the \textit{graph mode} in most cases and relies mostly on the \ac{KG} to do recommendations.
We also notice that with 100\% ratio, Recall is improved largely on ``Food-Kitchen'' dataset while MRR is not.
We think this is because of the density of the \ac{KG}.
As shown in Table~\ref{dataset2}, there are more triples in the `Movie-Book'' dataset, especially the ``Book'' domain.
The richer knowledge makes it relatively easier to rank the ground truth items.

Currently, using Algorithm \ref{alg:Extractgraph}, the ratios of the ground truth items in the \ac{KG} of the ``Movie-Book'' and ``Food-Kitchen'' datasets are 14\% and 12\%, respectively, both of which are relatively low.
Algorithm \ref{alg:Extractgraph} is only based on the entity distance calculated using the pretrained entity representations in the \ac{KG}, which is insufficient.
To this end, we think an important future research direction in order to further improve \ac{MIFN} is to design a more effective knowledge graph construction algorithm.

\subsection{Qualitative analysis with case studies (RQ4)}
\label{case_study}

\begin{figure}[h]
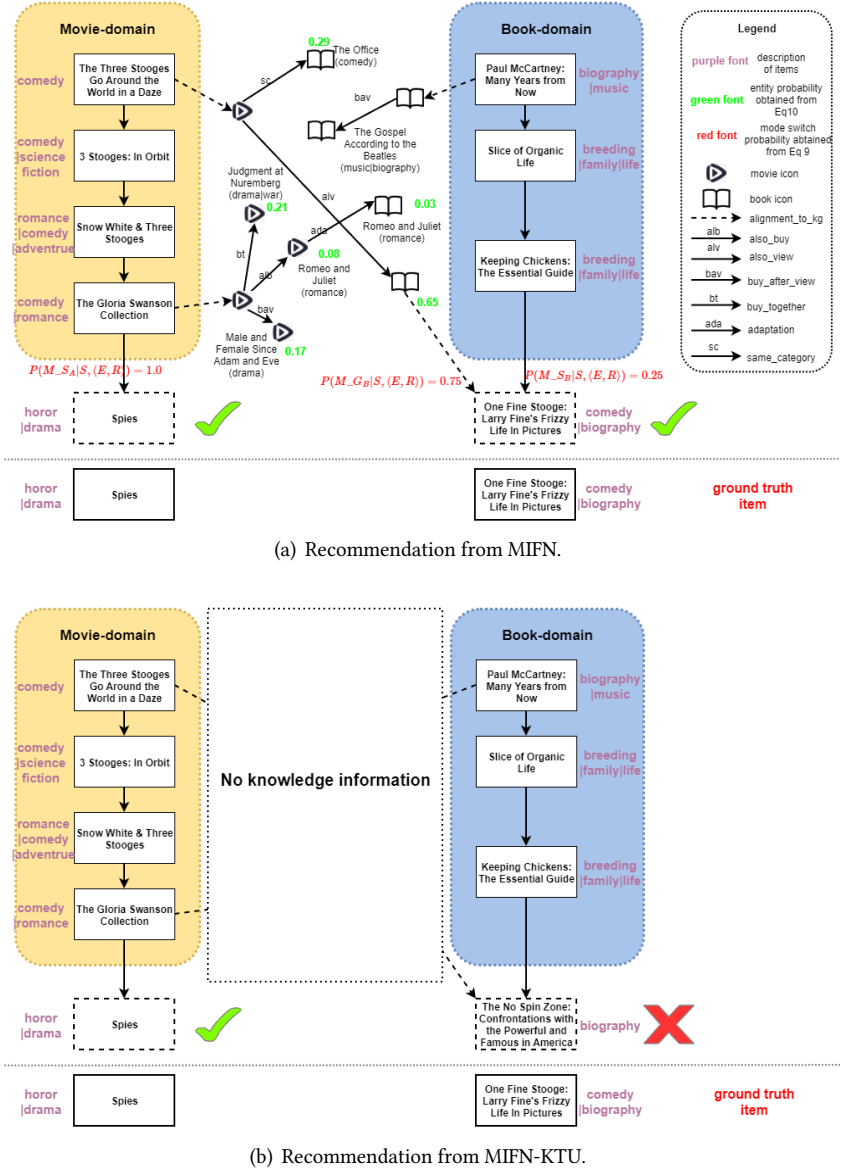

\centering
\subfigure[Recommendation from \ac{MIFN}.]{
\label{fig:1}
\includegraphics[clip,trim=7mm 2mm 7mm 7mm,width=0.8\textwidth]{figures/casestudy-MB1.png}}
\subfigure[Recommendation from \ac{MIFN}-\ac{KTU}.]{
\label{fig:2}
\includegraphics[clip,trim=7mm 2mm 7mm 7mm,width=0.8\textwidth]{figures/casestudy-MB1-noK.png}}
\caption{Case study of \ac{MIFN} and \ac{MIFN}-\ac{KTU} when the extracted \ac{KG} is relevant to the current user preference.}
\label{cs1}
\end{figure}

To analyze the recommendation results with and without a flow of knowledge, we list some examples from the ``Movie-Book'' dataset.
Figure~\ref{cs1} shows recommendations when the extracted \ac{KG} is relevant to the current user preference, and Figure~\ref{cs2} shows recommendations when the extracted \ac{KG} is irrelevant.
Figure~\ref{fig:1} and \ref{fig:3} are recommendations from \ac{MIFN}, while Figure~\ref{fig:2} and \ref{fig:4} are recommendations from \ac{MIFN}-\ac{KTU}.
In each figure, the orange color represents the interactions in the ``Movie'' domain, and the blue color represents the interactions in the ``Book'' domain.
The meaning of the different colored fonts and lines are explained in the legend.
The green tick indicates that the recommendation is correct, and the red cross indicates it is wrong.

\begin{figure}[h]
\centering
\subfigure[Recommendation from \ac{MIFN}.]{
\label{fig:3}
\includegraphics[clip,trim=7mm 2mm 7mm 7mm,width=0.8\textwidth]{figures/casestudy-MB2.png}}
\subfigure[Recommendation from \ac{MIFN}-\ac{KTU}.]{
\label{fig:4}
\includegraphics[clip,trim=7mm 2mm 7mm 7mm,width=0.8\textwidth]{figures/casestudy-MB2-noK.png}}
\caption{Case study of \ac{MIFN} and \ac{MIFN}-\ac{KTU} when the extracted \ac{KG} is irrelevant to the current user preference.}
\label{cs2}
\end{figure}

From Figure~\ref{cs1}, we can observe that when using the extracted \ac{KG} (Figure~\ref{fig:1}), \ac{MIFN} can give correct recommendations for both domains, however, the recommendation is wrong in the ``Book'' domain when the \ac{KG} is not used (Figure~\ref{fig:2}).
Furthermore, it should be noted that the mode switch probabilities in Figure~\ref{fig:1} are different for the two domains.
The probability of \textit{sequence mode} is 1.0 in the ``Movie'' domain, but in the ``Book'' domain, the probability of \textit{graph mode} is 0.75, which means that the recommendation of the ``Movie'' domain comes from the item set, while the recommendation mostly relies on the \ac{KG} for the ``Book'' domain.
From the \ac{KG}, the book entity ``One Fine Stooge: Larry Fine's Frizzy Life In Pictures'' gets the highest recommendation score of 0.65 in the ``Book'' domain.
The reason is that there exists a knowledge triple, that is, people who watch the movie ``The Three Stooges Go Around the World in a Daze'' will also view the book ``One Fine Stooge: Larry Fine's Frizzy Life In Pictures.''
This knowledge is well transferred by the flow of knowledge to obtain a better recommendation for the ``Book'' domain.
In contrast, as shown in Figure~\ref{fig:2}, the recommendation is wrong because the model only relies on the flow of behavioral information to recommend items from the item set.

On the other hand, we also found that the flow of knowledge is not always helpful.
As shown in Figure~\ref{fig:3}, \ac{MIFN} still gives a wrong recommendation even by modeling knowledge information flow with \ac{KTU}.
\ac{MIFN} recommends ``The confucian Transformation of Korea'' with the \textit{graph mode} recommendation probability of 0.77, which means, in this case, it still relies mostly on the \ac{KG} to do recommendation.
However, we can observe that most of the extracted entities in the \ac{KG} are thriller movies or reference books which are not relevant to the current user preference.
As a result, \ac{MIFN} performs somewhat worse than \ac{MIFN}-\ac{KTU}, because although \ac{MIFN}-\ac{KTU} also recommends the wrong item, the recommended item seems more relevant to the user preference, i.e., ``children|family.''
This suggests that the quality of the extracted \ac{KG} has a large impact on the final recommendation performance, which further verifies the conclusion in Section~\ref{algorithm_influence}.


\section{Conclusion and Future Work}
In this paper, we study how to incorporate knowledge into the \acl{CDSR} task.
We present \ac{MIFN}, which jointly models two types of information flow across domains, i.e., of behavioral information and of knowledge.
To verify the effectiveness of \ac{MIFN}, we conduct experiments on datasets from four Amazon domains.
The results demonstrate that \ac{MIFN} outperforms other state-of-the-art baselines.
Through extensive analysis experiments, we confirm that the flow of knowledge helps improve the recommendation performance in general, which means mixed flow of information can be used to enhance the \ac{CDSR} performance.


As to future work, \ac{MIFN} can be enhanced from at least two directions. 
First, currently we extract \acp{KG} with Algorithm~\ref{alg:Extractgraph} currently, which we show can be made more effective by including as much relevant knowledge as possible.
Hence, we will further study the knowledge graph construction algorithm to improve the quality of the \ac{KG} without increasing its size.
Second, \ac{MIFN} is limited to information flow between two domains in this work.
Therefore, we want to study how to make cross-domain recommendations across multiple domains.

\section*{Data and Code}
To facilitate reproduction of the results in the paper, we are sharing the code and resources used to produce our results at \url{https://github.com/mamuyang/MIFN}.

\begin{acks}
This research was partially supported by
the Natural Science Foundation of China (61972234, 61902219, 61672324, 61672322, 62072279), 
the Key Scientific and Technological Innovation Program of Shandong Province (2019JZZY010129), 
the Tencent AI Lab Rhino-Bird Focused Research Program (JR201932), 
the Fundamental Research Funds of Shandong University,
the National Key R\&D Program of China (2020YFB1406700),
and the Innovation Center for AI (ICAI).
All content represents the opinion of the authors, which is not necessarily shared or endorsed by their respective employers and/or sponsors.
\end{acks}

\bibliographystyle{ACM-Reference-Format}
\bibliography{bibtex.bib}


\begin{thebibliography}{80}


\ifx \showCODEN    \undefined \def \showCODEN     #1{\unskip}     \fi
\ifx \showDOI      \undefined \def \showDOI       #1{#1}\fi
\ifx \showISBNx    \undefined \def \showISBNx     #1{\unskip}     \fi
\ifx \showISBNxiii \undefined \def \showISBNxiii  #1{\unskip}     \fi
\ifx \showISSN     \undefined \def \showISSN      #1{\unskip}     \fi
\ifx \showLCCN     \undefined \def \showLCCN      #1{\unskip}     \fi
\ifx \shownote     \undefined \def \shownote      #1{#1}          \fi
\ifx \showarticletitle \undefined \def \showarticletitle #1{#1}   \fi
\ifx \showURL      \undefined \def \showURL       {\relax}        \fi
\providecommand\bibfield[2]{#2}
\providecommand\bibinfo[2]{#2}
\providecommand\natexlab[1]{#1}
\providecommand\showeprint[2][]{arXiv:#2}

\bibitem[\protect\citeauthoryear{Abel, Herder, Houben, Henze, and Krause}{Abel
  et~al\mbox{.}}{2013}]%
        {crosssys-usermodel}
\bibfield{author}{\bibinfo{person}{Fabian Abel}, \bibinfo{person}{Eelco
  Herder}, \bibinfo{person}{Geert-Jan Houben}, \bibinfo{person}{Nicola Henze},
  {and} \bibinfo{person}{Daniel Krause}.} \bibinfo{year}{2013}\natexlab{}.
\newblock \showarticletitle{Cross-system user modeling and personalization on
  the social web}.
\newblock \bibinfo{journal}{\emph{UMUAI 2013}}  \bibinfo{volume}{23}
  (\bibinfo{year}{2013}), \bibinfo{pages}{169--209}.
\newblock


\bibitem[\protect\citeauthoryear{Ai, Azizi, Chen, and Zhang}{Ai
  et~al\mbox{.}}{2018}]%
        {ai2018learning}
\bibfield{author}{\bibinfo{person}{Qingyao Ai}, \bibinfo{person}{Vahid Azizi},
  \bibinfo{person}{Xu Chen}, {and} \bibinfo{person}{Yongfeng Zhang}.}
  \bibinfo{year}{2018}\natexlab{}.
\newblock \showarticletitle{Learning heterogeneous knowledge base embeddings
  for explainable recommendation}.
\newblock \bibinfo{journal}{\emph{Algorithms 2018}}  \bibinfo{volume}{11}
  (\bibinfo{year}{2018}), \bibinfo{pages}{137}.
\newblock


\bibitem[\protect\citeauthoryear{Berkovsky, Kuflik, and Ricci}{Berkovsky
  et~al\mbox{.}}{2007}]%
        {BerkovskyKR07}
\bibfield{author}{\bibinfo{person}{Shlomo Berkovsky}, \bibinfo{person}{Tsvi
  Kuflik}, {and} \bibinfo{person}{Francesco Ricci}.}
  \bibinfo{year}{2007}\natexlab{}.
\newblock \showarticletitle{Distributed collaborative filtering with domain
  specialization}. In \bibinfo{booktitle}{\emph{RecSys 2007}}.
  \bibinfo{pages}{33--40}.
\newblock


\bibitem[\protect\citeauthoryear{Berkovsky, Kuflik, and Ricci}{Berkovsky
  et~al\mbox{.}}{2008}]%
        {enhance-person}
\bibfield{author}{\bibinfo{person}{Shlomo Berkovsky}, \bibinfo{person}{Tsvi
  Kuflik}, {and} \bibinfo{person}{Francesco Ricci}.}
  \bibinfo{year}{2008}\natexlab{}.
\newblock \showarticletitle{Mediation of user models for enhanced
  personalization in recommender systems}.
\newblock \bibinfo{journal}{\emph{UMUAI 2008}}  \bibinfo{volume}{18}
  (\bibinfo{year}{2008}), \bibinfo{pages}{245--286}.
\newblock


\bibitem[\protect\citeauthoryear{Bi, Song, Yao, Wu, Wang, and Xiao}{Bi
  et~al\mbox{.}}{2020}]%
        {ye2020dcdir}
\bibfield{author}{\bibinfo{person}{Ye Bi}, \bibinfo{person}{Liqiang Song},
  \bibinfo{person}{Mengqiu Yao}, \bibinfo{person}{Zhenyu Wu},
  \bibinfo{person}{Jianming Wang}, {and} \bibinfo{person}{Jing Xiao}.}
  \bibinfo{year}{2020}\natexlab{}.
\newblock \showarticletitle{DCDIR: A deep cross-domain recommendation system
  for cold start users in insurance domain}. In \bibinfo{booktitle}{\emph{SIGIR
  2020}}. \bibinfo{pages}{1661--1664}.
\newblock


\bibitem[\protect\citeauthoryear{Bogina and Kuflik}{Bogina and Kuflik}{2017}]%
        {dwelltimernn}
\bibfield{author}{\bibinfo{person}{Veronika Bogina} {and} \bibinfo{person}{Tsvi
  Kuflik}.} \bibinfo{year}{2017}\natexlab{}.
\newblock \showarticletitle{Incorporating dwell time in session-based
  recommendations with recurrent Neural networks}. In
  \bibinfo{booktitle}{\emph{RecSys 2017}}. \bibinfo{pages}{57--59}.
\newblock


\bibitem[\protect\citeauthoryear{Chen, Moore, Turnbull, and Joachims}{Chen
  et~al\mbox{.}}{2012}]%
        {chen2012playlist}
\bibfield{author}{\bibinfo{person}{Shuo Chen}, \bibinfo{person}{Josh~L Moore},
  \bibinfo{person}{Douglas Turnbull}, {and} \bibinfo{person}{Thorsten
  Joachims}.} \bibinfo{year}{2012}\natexlab{}.
\newblock \showarticletitle{Playlist prediction via metric embedding}. In
  \bibinfo{booktitle}{\emph{SIGKDD 2012}}. \bibinfo{pages}{714--722}.
\newblock


\bibitem[\protect\citeauthoryear{Chen, Xu, Zhang, Tang, Cao, Qin, and Zha}{Chen
  et~al\mbox{.}}{2018}]%
        {chen2018sequential_memNet}
\bibfield{author}{\bibinfo{person}{Xu Chen}, \bibinfo{person}{Hongteng Xu},
  \bibinfo{person}{Yongfeng Zhang}, \bibinfo{person}{Jiaxi Tang},
  \bibinfo{person}{Yixin Cao}, \bibinfo{person}{Zheng Qin}, {and}
  \bibinfo{person}{Hongyuan Zha}.} \bibinfo{year}{2018}\natexlab{}.
\newblock \showarticletitle{Sequential recommendation with user memory
  networks}. In \bibinfo{booktitle}{\emph{WSDM 2018}}.
  \bibinfo{pages}{108--116}.
\newblock


\bibitem[\protect\citeauthoryear{Cheng, Ding, Zhu, and Kankanhalli}{Cheng
  et~al\mbox{.}}{2018}]%
        {latentfactormodel}
\bibfield{author}{\bibinfo{person}{Zhiyong Cheng}, \bibinfo{person}{Ying Ding},
  \bibinfo{person}{Lei Zhu}, {and} \bibinfo{person}{Mohan~S. Kankanhalli}.}
  \bibinfo{year}{2018}\natexlab{}.
\newblock \showarticletitle{Aspect-aware latent factor model: Rating prediction
  with ratings and reviews}. In \bibinfo{booktitle}{\emph{WWW 2018}}.
  \bibinfo{pages}{639--648}.
\newblock


\bibitem[\protect\citeauthoryear{Cremonesi, Tripodi, and Turrin}{Cremonesi
  et~al\mbox{.}}{2011}]%
        {CremonesiTT11}
\bibfield{author}{\bibinfo{person}{Paolo Cremonesi}, \bibinfo{person}{Antonio
  Tripodi}, {and} \bibinfo{person}{Roberto Turrin}.}
  \bibinfo{year}{2011}\natexlab{}.
\newblock \showarticletitle{Cross-domain recommender systems}. In
  \bibinfo{booktitle}{\emph{ICDMW 2011}}. \bibinfo{pages}{496--503}.
\newblock


\bibitem[\protect\citeauthoryear{Donkers, Loepp, and Ziegler}{Donkers
  et~al\mbox{.}}{2017}]%
        {donkers2017sequential}
\bibfield{author}{\bibinfo{person}{Tim Donkers}, \bibinfo{person}{Benedikt
  Loepp}, {and} \bibinfo{person}{J{\"u}rgen Ziegler}.}
  \bibinfo{year}{2017}\natexlab{}.
\newblock \showarticletitle{Sequential user-based recurrent neural network
  recommendations}. In \bibinfo{booktitle}{\emph{RecSys 2017}}.
  \bibinfo{pages}{152--160}.
\newblock


\bibitem[\protect\citeauthoryear{Elkahky, Song, and He}{Elkahky
  et~al\mbox{.}}{2015}]%
        {DSSM2015}
\bibfield{author}{\bibinfo{person}{Ali~Mamdouh Elkahky}, \bibinfo{person}{Yang
  Song}, {and} \bibinfo{person}{Xiaodong He}.} \bibinfo{year}{2015}\natexlab{}.
\newblock \showarticletitle{A multi-view deep learning approach for cross
  domain user modeling in recommendation systems}. In
  \bibinfo{booktitle}{\emph{WWW 2015}}. \bibinfo{pages}{278--288}.
\newblock


\bibitem[\protect\citeauthoryear{Fern{\'a}ndez-Tob{\'\i}as, Cantador,
  Kaminskas, and Ricci}{Fern{\'a}ndez-Tob{\'\i}as et~al\mbox{.}}{2012}]%
        {cd_survey}
\bibfield{author}{\bibinfo{person}{Ignacio Fern{\'a}ndez-Tob{\'\i}as},
  \bibinfo{person}{Iv{\'a}n Cantador}, \bibinfo{person}{Marius Kaminskas},
  {and} \bibinfo{person}{Francesco Ricci}.} \bibinfo{year}{2012}\natexlab{}.
\newblock \showarticletitle{Cross-domain recommender systems: A survey of the
  state of the art}. In \bibinfo{booktitle}{\emph{CERI 2012}}.
  \bibinfo{pages}{24}.
\newblock


\bibitem[\protect\citeauthoryear{Fern{\'{a}}ndez{-}Tob{\'{\i}}as, Cantador, and
  Plaza}{Fern{\'{a}}ndez{-}Tob{\'{\i}}as et~al\mbox{.}}{2013}]%
        {Fernandez-TobiasCP13}
\bibfield{author}{\bibinfo{person}{Ignacio Fern{\'{a}}ndez{-}Tob{\'{\i}}as},
  \bibinfo{person}{Iv{\'{a}}n Cantador}, {and} \bibinfo{person}{Laura Plaza}.}
  \bibinfo{year}{2013}\natexlab{}.
\newblock \showarticletitle{An emotion dimensional model based on social tags:
  Crossing folksonomies and enhancing recommendations}. In
  \bibinfo{booktitle}{\emph{EC-Web 2013}}. \bibinfo{pages}{88--100}.
\newblock


\bibitem[\protect\citeauthoryear{Fern{\'a}ndez-Tob{\'\i}as, Cantador, Tomeo,
  Anelli, and Di~Noia}{Fern{\'a}ndez-Tob{\'\i}as et~al\mbox{.}}{2019}]%
        {fernandez2019addressing}
\bibfield{author}{\bibinfo{person}{Ignacio Fern{\'a}ndez-Tob{\'\i}as},
  \bibinfo{person}{Iv{\'a}n Cantador}, \bibinfo{person}{Paolo Tomeo},
  \bibinfo{person}{Vito~Walter Anelli}, {and} \bibinfo{person}{Tommaso
  Di~Noia}.} \bibinfo{year}{2019}\natexlab{}.
\newblock \showarticletitle{Addressing the user cold start with cross-domain
  collaborative filtering: Exploiting item metadata in matrix factorization}.
\newblock \bibinfo{journal}{\emph{UMUAI 2019}}  \bibinfo{volume}{29}
  (\bibinfo{year}{2019}), \bibinfo{pages}{443--486}.
\newblock


\bibitem[\protect\citeauthoryear{Gao, Chen, Feng, Zhao, He, Li, and Jin}{Gao
  et~al\mbox{.}}{2019}]%
        {gao2019cross}
\bibfield{author}{\bibinfo{person}{Chen Gao}, \bibinfo{person}{Xiangning Chen},
  \bibinfo{person}{Fuli Feng}, \bibinfo{person}{Kai Zhao},
  \bibinfo{person}{Xiangnan He}, \bibinfo{person}{Yong Li}, {and}
  \bibinfo{person}{Depeng Jin}.} \bibinfo{year}{2019}\natexlab{}.
\newblock \showarticletitle{Cross-domain recommendation without sharing
  user-relevant data}. In \bibinfo{booktitle}{\emph{WWW 2019}}.
  \bibinfo{pages}{491--502}.
\newblock


\bibitem[\protect\citeauthoryear{Givon and Lavrenko}{Givon and
  Lavrenko}{2009}]%
        {Givon2009}
\bibfield{author}{\bibinfo{person}{Sharon Givon} {and} \bibinfo{person}{Victor
  Lavrenko}.} \bibinfo{year}{2009}\natexlab{}.
\newblock \showarticletitle{Predicting social-tags for cold start book
  recommendations}. In \bibinfo{booktitle}{\emph{RecSys 2009}}.
  \bibinfo{pages}{333--336}.
\newblock


\bibitem[\protect\citeauthoryear{He and McAuley}{He and McAuley}{2016}]%
        {he2016fusing}
\bibfield{author}{\bibinfo{person}{Ruining He} {and} \bibinfo{person}{Julian
  McAuley}.} \bibinfo{year}{2016}\natexlab{}.
\newblock \showarticletitle{Fusing similarity models with markov chains for
  sparse sequential recommendation}. In \bibinfo{booktitle}{\emph{ICDM 2016}}.
  \bibinfo{pages}{191--200}.
\newblock


\bibitem[\protect\citeauthoryear{He, He, Du, and Chua}{He
  et~al\mbox{.}}{2018}]%
        {apranking}
\bibfield{author}{\bibinfo{person}{Xiangnan He}, \bibinfo{person}{Zhankui He},
  \bibinfo{person}{Xiaoyu Du}, {and} \bibinfo{person}{Tat-Seng Chua}.}
  \bibinfo{year}{2018}\natexlab{}.
\newblock \showarticletitle{Adversarial personalized ranking for
  recommendation}. In \bibinfo{booktitle}{\emph{SIGIR 2018}}.
  \bibinfo{pages}{355--364}.
\newblock


\bibitem[\protect\citeauthoryear{He, Liao, Zhang, Nie, Hu, and Chua}{He
  et~al\mbox{.}}{2017}]%
        {ncf_he}
\bibfield{author}{\bibinfo{person}{Xiangnan He}, \bibinfo{person}{Lizi Liao},
  \bibinfo{person}{Hanwang Zhang}, \bibinfo{person}{Liqiang Nie},
  \bibinfo{person}{Xia Hu}, {and} \bibinfo{person}{Tat-Seng Chua}.}
  \bibinfo{year}{2017}\natexlab{}.
\newblock \showarticletitle{Neural collaborative filtering}. In
  \bibinfo{booktitle}{\emph{WWW 2017}}. \bibinfo{pages}{173--182}.
\newblock


\bibitem[\protect\citeauthoryear{Hidasi, Karatzoglou, Baltrunas, and
  Tikk}{Hidasi et~al\mbox{.}}{2016a}]%
        {GRU4rec}
\bibfield{author}{\bibinfo{person}{Bal{\'{a}}zs Hidasi},
  \bibinfo{person}{Alexandros Karatzoglou}, \bibinfo{person}{Linas Baltrunas},
  {and} \bibinfo{person}{Domonkos Tikk}.} \bibinfo{year}{2016}\natexlab{a}.
\newblock \showarticletitle{Session-based recommendations with recurrent neural
  networks}. In \bibinfo{booktitle}{\emph{ICLR 2016}}. \bibinfo{pages}{--}.
\newblock


\bibitem[\protect\citeauthoryear{Hidasi, Quadrana, Karatzoglou, and
  Tikk}{Hidasi et~al\mbox{.}}{2016b}]%
        {hidasi2016parallel}
\bibfield{author}{\bibinfo{person}{Bal{\'a}zs Hidasi}, \bibinfo{person}{Massimo
  Quadrana}, \bibinfo{person}{Alexandros Karatzoglou}, {and}
  \bibinfo{person}{Domonkos Tikk}.} \bibinfo{year}{2016}\natexlab{b}.
\newblock \showarticletitle{Parallel recurrent neural network architectures for
  feature-rich session-based recommendations}. In
  \bibinfo{booktitle}{\emph{RecSys 2016}}. \bibinfo{pages}{241--248}.
\newblock


\bibitem[\protect\citeauthoryear{Hu, Zhang, and Yang}{Hu
  et~al\mbox{.}}{2018b}]%
        {Conet}
\bibfield{author}{\bibinfo{person}{Guangneng Hu}, \bibinfo{person}{Yu Zhang},
  {and} \bibinfo{person}{Qiang Yang}.} \bibinfo{year}{2018}\natexlab{b}.
\newblock \showarticletitle{CoNet: Collaborative cross networks for
  cross-domain recommendation}. In \bibinfo{booktitle}{\emph{CIKM 2018}}.
  \bibinfo{pages}{667--676}.
\newblock


\bibitem[\protect\citeauthoryear{Hu, Dai, Qiu, Xia, Li, Huang, and Chen}{Hu
  et~al\mbox{.}}{2018a}]%
        {huCF2018tkdd}
\bibfield{author}{\bibinfo{person}{Guang-Neng Hu}, \bibinfo{person}{Xin-Yu
  Dai}, \bibinfo{person}{Feng-Yu Qiu}, \bibinfo{person}{Rui Xia},
  \bibinfo{person}{Tao Li}, \bibinfo{person}{Shu-Jian Huang}, {and}
  \bibinfo{person}{Jia-Jun Chen}.} \bibinfo{year}{2018}\natexlab{a}.
\newblock \showarticletitle{Collaborative filtering with topic and social
  latent factors incorporating implicit feedback}.
\newblock \bibinfo{journal}{\emph{TKDD 2018}}  \bibinfo{volume}{12}
  (\bibinfo{year}{2018}), \bibinfo{pages}{1--30}.
\newblock


\bibitem[\protect\citeauthoryear{Hu, Cao, Xu, Cao, Gu, and Zhu}{Hu
  et~al\mbox{.}}{2013}]%
        {hu2013}
\bibfield{author}{\bibinfo{person}{Liang Hu}, \bibinfo{person}{Jian Cao},
  \bibinfo{person}{Guandong Xu}, \bibinfo{person}{Longbing Cao},
  \bibinfo{person}{Zhiping Gu}, {and} \bibinfo{person}{Can Zhu}.}
  \bibinfo{year}{2013}\natexlab{}.
\newblock \showarticletitle{Personalized recommendation via cross-domain
  triadic factorization}. In \bibinfo{booktitle}{\emph{WWW 2013}}.
  \bibinfo{pages}{595--606}.
\newblock


\bibitem[\protect\citeauthoryear{Huang, Ren, Zhao, He, Wen, and Dong}{Huang
  et~al\mbox{.}}{2019}]%
        {huang2019taxonomy}
\bibfield{author}{\bibinfo{person}{Jin Huang}, \bibinfo{person}{Zhaochun Ren},
  \bibinfo{person}{Wayne~Xin Zhao}, \bibinfo{person}{Gaole He},
  \bibinfo{person}{Ji-Rong Wen}, {and} \bibinfo{person}{Daxiang Dong}.}
  \bibinfo{year}{2019}\natexlab{}.
\newblock \showarticletitle{Taxonomy-aware multi-hop reasoning networks for
  sequential recommendation}. In \bibinfo{booktitle}{\emph{WSDM 2019}}.
  \bibinfo{pages}{573--581}.
\newblock


\bibitem[\protect\citeauthoryear{Huang, Zhao, Dou, Wen, and Chang}{Huang
  et~al\mbox{.}}{2018}]%
        {KSR}
\bibfield{author}{\bibinfo{person}{Jin Huang}, \bibinfo{person}{Wayne~Xin
  Zhao}, \bibinfo{person}{Hongjian Dou}, \bibinfo{person}{Ji-Rong Wen}, {and}
  \bibinfo{person}{Edward~Y Chang}.} \bibinfo{year}{2018}\natexlab{}.
\newblock \showarticletitle{Improving sequential recommendation with
  knowledge-enhanced memory networks}. In \bibinfo{booktitle}{\emph{SIGIR
  2018}}. \bibinfo{pages}{505--514}.
\newblock


\bibitem[\protect\citeauthoryear{Kipf and Welling}{Kipf and Welling}{2017}]%
        {kipf2016semi}
\bibfield{author}{\bibinfo{person}{Thomas~N Kipf} {and} \bibinfo{person}{Max
  Welling}.} \bibinfo{year}{2017}\natexlab{}.
\newblock \showarticletitle{Semi-supervised classification with graph
  convolutional networks}. In \bibinfo{booktitle}{\emph{ICLR 2017}}.
\newblock


\bibitem[\protect\citeauthoryear{Krishnan, Das, Bendre, Yang, and
  Sundaram}{Krishnan et~al\mbox{.}}{2020}]%
        {krishnan2020transfer}
\bibfield{author}{\bibinfo{person}{Adit Krishnan}, \bibinfo{person}{Mahashweta
  Das}, \bibinfo{person}{Mangesh Bendre}, \bibinfo{person}{Hao Yang}, {and}
  \bibinfo{person}{Hari Sundaram}.} \bibinfo{year}{2020}\natexlab{}.
\newblock \showarticletitle{Transfer learning via contextual invariants for
  one-to-many cross-domain recommendation}. In \bibinfo{booktitle}{\emph{SIGIR
  2020}}. \bibinfo{pages}{1081--1090}.
\newblock


\bibitem[\protect\citeauthoryear{Li, Yang, and Xue}{Li et~al\mbox{.}}{2009}]%
        {cd-cf}
\bibfield{author}{\bibinfo{person}{Bin Li}, \bibinfo{person}{Qiang Yang}, {and}
  \bibinfo{person}{Xiangyang Xue}.} \bibinfo{year}{2009}\natexlab{}.
\newblock \showarticletitle{Can movies and books collaborate? Cross-domain
  collaborative filtering for sparsity reduction}. In
  \bibinfo{booktitle}{\emph{IJCAI 2009}}. \bibinfo{pages}{2052--2057}.
\newblock


\bibitem[\protect\citeauthoryear{Li, Niu, Luo, Chen, and Quan}{Li
  et~al\mbox{.}}{2019}]%
        {review_driven_sr2019}
\bibfield{author}{\bibinfo{person}{Chenliang Li}, \bibinfo{person}{Xichuan
  Niu}, \bibinfo{person}{Xiangyang Luo}, \bibinfo{person}{Zhenzhong Chen},
  {and} \bibinfo{person}{Cong Quan}.} \bibinfo{year}{2019}\natexlab{}.
\newblock \showarticletitle{A review-driven neural model for sequential
  recommendation}. In \bibinfo{booktitle}{\emph{IJCAI 2019}}.
  \bibinfo{pages}{2866--2872}.
\newblock


\bibitem[\protect\citeauthoryear{Li, Ren, Chen, Ren, Lian, and Ma}{Li
  et~al\mbox{.}}{2017}]%
        {NARM}
\bibfield{author}{\bibinfo{person}{Jing Li}, \bibinfo{person}{Pengjie Ren},
  \bibinfo{person}{Zhumin Chen}, \bibinfo{person}{Zhaochun Ren},
  \bibinfo{person}{Tao Lian}, {and} \bibinfo{person}{Jun Ma}.}
  \bibinfo{year}{2017}\natexlab{}.
\newblock \showarticletitle{Neural attentive session-based recommendation}. In
  \bibinfo{booktitle}{\emph{CIKM 2017}}. \bibinfo{pages}{1419--1428}.
\newblock


\bibitem[\protect\citeauthoryear{Li and Tuzhilin}{Li and Tuzhilin}{2019}]%
        {li2019ddtcdr}
\bibfield{author}{\bibinfo{person}{Pan Li} {and} \bibinfo{person}{Alexander
  Tuzhilin}.} \bibinfo{year}{2019}\natexlab{}.
\newblock \showarticletitle{DDTCDR: Deep dual transfer cross domain
  recommendation}. In \bibinfo{booktitle}{\emph{WSDM 2020}}.
  \bibinfo{pages}{331--339}.
\newblock


\bibitem[\protect\citeauthoryear{Lian, Zhang, Xie, and Sun}{Lian
  et~al\mbox{.}}{2017}]%
        {CCCFNet}
\bibfield{author}{\bibinfo{person}{Jianxun Lian}, \bibinfo{person}{Fuzheng
  Zhang}, \bibinfo{person}{Xing Xie}, {and} \bibinfo{person}{Guangzhong Sun}.}
  \bibinfo{year}{2017}\natexlab{}.
\newblock \showarticletitle{CCCFNet: A content-boosted collaborative filtering
  neural network for cross domain recommender systems}. In
  \bibinfo{booktitle}{\emph{WWW 2017}}. \bibinfo{pages}{817--818}.
\newblock


\bibitem[\protect\citeauthoryear{Lin, Ren, Chen, Ren, Ma, and de~Rijke}{Lin
  et~al\mbox{.}}{2019}]%
        {lin2019explainable}
\bibfield{author}{\bibinfo{person}{Yujie Lin}, \bibinfo{person}{Pengjie Ren},
  \bibinfo{person}{Zhumin Chen}, \bibinfo{person}{Zhaochun Ren},
  \bibinfo{person}{Jun Ma}, {and} \bibinfo{person}{Maarten de Rijke}.}
  \bibinfo{year}{2019}\natexlab{}.
\newblock \showarticletitle{Explainable outfit recommendation with joint outfit
  matching and comment generation}.
\newblock \bibinfo{journal}{\emph{TKDE 2019}} \bibinfo{volume}{32},
  \bibinfo{number}{8} (\bibinfo{year}{2019}), \bibinfo{pages}{1502--1516}.
\newblock


\bibitem[\protect\citeauthoryear{Liu, Wu, Wang, Li, and Wang}{Liu
  et~al\mbox{.}}{2016}]%
        {liu2016context}
\bibfield{author}{\bibinfo{person}{Qiang Liu}, \bibinfo{person}{Shu Wu},
  \bibinfo{person}{Diyi Wang}, \bibinfo{person}{Zhaokang Li}, {and}
  \bibinfo{person}{Liang Wang}.} \bibinfo{year}{2016}\natexlab{}.
\newblock \showarticletitle{Context-aware sequential recommendation}. In
  \bibinfo{booktitle}{\emph{ICDM 2016}}. \bibinfo{pages}{1053--1058}.
\newblock


\bibitem[\protect\citeauthoryear{Liu, Zeng, Mokhosi, and Zhang}{Liu
  et~al\mbox{.}}{2018}]%
        {STAMP}
\bibfield{author}{\bibinfo{person}{Qiao Liu}, \bibinfo{person}{Yifu Zeng},
  \bibinfo{person}{Refuoe Mokhosi}, {and} \bibinfo{person}{Haibin Zhang}.}
  \bibinfo{year}{2018}\natexlab{}.
\newblock \showarticletitle{STAMP: Short-term attention/memory priority model
  for session-based recommendation}. In \bibinfo{booktitle}{\emph{SIGKDD
  2018}}. \bibinfo{pages}{1831--1839}.
\newblock


\bibitem[\protect\citeauthoryear{Loni, Shi, Larson, and Hanjalic}{Loni
  et~al\mbox{.}}{2014}]%
        {LoniSLH14}
\bibfield{author}{\bibinfo{person}{Babak Loni}, \bibinfo{person}{Yue Shi},
  \bibinfo{person}{Martha~A. Larson}, {and} \bibinfo{person}{Alan Hanjalic}.}
  \bibinfo{year}{2014}\natexlab{}.
\newblock \showarticletitle{Cross-domain collaborative filtering with
  factorization machines}. In \bibinfo{booktitle}{\emph{ECIR 2014}}.
  \bibinfo{pages}{656--661}.
\newblock


\bibitem[\protect\citeauthoryear{Ma, Ren, Lin, Chen, Ma, and de~Rijke}{Ma
  et~al\mbox{.}}{2019a}]%
        {ma2019pi}
\bibfield{author}{\bibinfo{person}{Muyang Ma}, \bibinfo{person}{Pengjie Ren},
  \bibinfo{person}{Yujie Lin}, \bibinfo{person}{Zhumin Chen},
  \bibinfo{person}{Jun Ma}, {and} \bibinfo{person}{Maarten de Rijke}.}
  \bibinfo{year}{2019}\natexlab{a}.
\newblock \showarticletitle{$\pi$-Net: A Parallel information-sharing network
  for shared-account cross-domain sequential recommendations}. In
  \bibinfo{booktitle}{\emph{SIGIR 2019}}. \bibinfo{pages}{685--694}.
\newblock


\bibitem[\protect\citeauthoryear{Ma, Zhang, Wang, Cui, and Huang}{Ma
  et~al\mbox{.}}{2018a}]%
        {ma2018mention}
\bibfield{author}{\bibinfo{person}{Renfeng Ma}, \bibinfo{person}{Qi Zhang},
  \bibinfo{person}{Jiawen Wang}, \bibinfo{person}{Lizhen Cui}, {and}
  \bibinfo{person}{Xuanjing Huang}.} \bibinfo{year}{2018}\natexlab{a}.
\newblock \showarticletitle{Mention recommendation for multimodal microblog
  with cross-attention memory network}. In \bibinfo{booktitle}{\emph{SIGIR
  2018}}. \bibinfo{pages}{195--204}.
\newblock


\bibitem[\protect\citeauthoryear{Ma, Zhang, Cao, Jin, Wang, Liu, Ma, and
  Ren}{Ma et~al\mbox{.}}{2019b}]%
        {ma2019jointly}
\bibfield{author}{\bibinfo{person}{Weizhi Ma}, \bibinfo{person}{Min Zhang},
  \bibinfo{person}{Yue Cao}, \bibinfo{person}{Woojeong Jin},
  \bibinfo{person}{Chenyang Wang}, \bibinfo{person}{Yiqun Liu},
  \bibinfo{person}{Shaoping Ma}, {and} \bibinfo{person}{Xiang Ren}.}
  \bibinfo{year}{2019}\natexlab{b}.
\newblock \showarticletitle{Jointly learning explainable rules for
  recommendation with knowledge graph}. In \bibinfo{booktitle}{\emph{WWW
  2019}}. \bibinfo{pages}{1210--1221}.
\newblock


\bibitem[\protect\citeauthoryear{Ma, Zhang, Wang, Luo, Liu, and Ma}{Ma
  et~al\mbox{.}}{2018b}]%
        {MaZWLLM18}
\bibfield{author}{\bibinfo{person}{Weizhi Ma}, \bibinfo{person}{Min Zhang},
  \bibinfo{person}{Chenyang Wang}, \bibinfo{person}{Cheng Luo},
  \bibinfo{person}{Yiqun Liu}, {and} \bibinfo{person}{Shaoping Ma}.}
  \bibinfo{year}{2018}\natexlab{b}.
\newblock \showarticletitle{Your tweets reveal what you like: Introducing
  cross-media content information into multi-domain recommendation}. In
  \bibinfo{booktitle}{\emph{IJCAI 2018}}. \bibinfo{pages}{3484--3490}.
\newblock


\bibitem[\protect\citeauthoryear{Mei, Ren, Chen, Nie, Ma, and Nie}{Mei
  et~al\mbox{.}}{2018}]%
        {attnetwork_mei}
\bibfield{author}{\bibinfo{person}{Lei Mei}, \bibinfo{person}{Pengjie Ren},
  \bibinfo{person}{Zhumin Chen}, \bibinfo{person}{Liqiang Nie},
  \bibinfo{person}{Jun Ma}, {and} \bibinfo{person}{Jian-Yun Nie}.}
  \bibinfo{year}{2018}\natexlab{}.
\newblock \showarticletitle{An attentive interaction network for context-aware
  recommendations}. In \bibinfo{booktitle}{\emph{CIKM 2018}}.
  \bibinfo{pages}{157--166}.
\newblock


\bibitem[\protect\citeauthoryear{Mirbakhsh and Ling}{Mirbakhsh and
  Ling}{2015}]%
        {improving2015tkdd}
\bibfield{author}{\bibinfo{person}{Nima Mirbakhsh} {and}
  \bibinfo{person}{Charles~X Ling}.} \bibinfo{year}{2015}\natexlab{}.
\newblock \showarticletitle{Improving top-n recommendation for cold-start users
  via cross-domain information}.
\newblock \bibinfo{journal}{\emph{TKDD 2015}}  \bibinfo{volume}{9}
  (\bibinfo{year}{2015}), \bibinfo{pages}{1--19}.
\newblock


\bibitem[\protect\citeauthoryear{Misra, Shrivastava, Gupta, and Hebert}{Misra
  et~al\mbox{.}}{2016}]%
        {misra2016cross}
\bibfield{author}{\bibinfo{person}{Ishan Misra}, \bibinfo{person}{Abhinav
  Shrivastava}, \bibinfo{person}{Abhinav Gupta}, {and} \bibinfo{person}{Martial
  Hebert}.} \bibinfo{year}{2016}\natexlab{}.
\newblock \showarticletitle{Cross-stitch networks for multi-task learning}. In
  \bibinfo{booktitle}{\emph{CVPR 2016}}. \bibinfo{pages}{3994--4003}.
\newblock


\bibitem[\protect\citeauthoryear{Pan, Xiang, Liu, and Yang}{Pan
  et~al\mbox{.}}{2010}]%
        {transfer-cf}
\bibfield{author}{\bibinfo{person}{Weike Pan}, \bibinfo{person}{Evan~Wei
  Xiang}, \bibinfo{person}{Nathan~Nan Liu}, {and} \bibinfo{person}{Qiang
  Yang}.} \bibinfo{year}{2010}\natexlab{}.
\newblock \showarticletitle{Transfer learning in collaborative filtering for
  sparsity reduction}. In \bibinfo{booktitle}{\emph{AAAI 2010}}.
  \bibinfo{pages}{230--235}.
\newblock


\bibitem[\protect\citeauthoryear{Quadrana, Cremonesi, and Jannach}{Quadrana
  et~al\mbox{.}}{2018}]%
        {quadrana2018sequence}
\bibfield{author}{\bibinfo{person}{Massimo Quadrana}, \bibinfo{person}{Paolo
  Cremonesi}, {and} \bibinfo{person}{Dietmar Jannach}.}
  \bibinfo{year}{2018}\natexlab{}.
\newblock \showarticletitle{Sequence-aware recommender systems}.
\newblock \bibinfo{journal}{\emph{CSUR 2018}}  \bibinfo{volume}{51}
  (\bibinfo{year}{2018}), \bibinfo{pages}{66}.
\newblock


\bibitem[\protect\citeauthoryear{Quadrana, Karatzoglou, Hidasi, and
  Cremonesi}{Quadrana et~al\mbox{.}}{2017}]%
        {HRNN}
\bibfield{author}{\bibinfo{person}{Massimo Quadrana},
  \bibinfo{person}{Alexandros Karatzoglou}, \bibinfo{person}{Balzs Hidasi},
  {and} \bibinfo{person}{Paolo Cremonesi}.} \bibinfo{year}{2017}\natexlab{}.
\newblock \showarticletitle{Personalizing session-based recommendations with
  hierarchical recurrent neural networks}. In \bibinfo{booktitle}{\emph{RecSys
  2017}}. \bibinfo{pages}{130--137}.
\newblock


\bibitem[\protect\citeauthoryear{Ren, Chen, Li, Ren, Ma, and de~Rijke}{Ren
  et~al\mbox{.}}{2019}]%
        {repeatnet}
\bibfield{author}{\bibinfo{person}{Pengjie Ren}, \bibinfo{person}{Zhumin Chen},
  \bibinfo{person}{Jing Li}, \bibinfo{person}{Zhaochun Ren},
  \bibinfo{person}{Jun Ma}, {and} \bibinfo{person}{Maarten de Rijke}.}
  \bibinfo{year}{2019}\natexlab{}.
\newblock \showarticletitle{RepeatNet: A repeat aware neural recommendation
  machine for session-based recommendation}. In \bibinfo{booktitle}{\emph{AAAI
  2019}}. \bibinfo{pages}{4806--4813}.
\newblock


\bibitem[\protect\citeauthoryear{Ren, Ren, Sun, He, Yin, and de~Rijke}{Ren
  et~al\mbox{.}}{2020b}]%
        {10.1145/3336191.3371884}
\bibfield{author}{\bibinfo{person}{Pengjie Ren}, \bibinfo{person}{Zhaochun
  Ren}, \bibinfo{person}{Fei Sun}, \bibinfo{person}{Xiangnan He},
  \bibinfo{person}{Dawei Yin}, {and} \bibinfo{person}{Maarten de Rijke}.}
  \bibinfo{year}{2020}\natexlab{b}.
\newblock \showarticletitle{NLP4REC: The WSDM 2020 workshop on natural language
  processing for recommendations}. In \bibinfo{booktitle}{\emph{WSDM 2020}}.
  \bibinfo{pages}{907--908}.
\newblock


\bibitem[\protect\citeauthoryear{Ren, Liu, Li, Zhao, Wang, Ding, and Wen}{Ren
  et~al\mbox{.}}{2020a}]%
        {masr2020}
\bibfield{author}{\bibinfo{person}{Ruiyang Ren}, \bibinfo{person}{Zhaoyang
  Liu}, \bibinfo{person}{Yaliang Li}, \bibinfo{person}{Wayne~Xin Zhao},
  \bibinfo{person}{Hui Wang}, \bibinfo{person}{Bolin Ding}, {and}
  \bibinfo{person}{Ji{-}Rong Wen}.} \bibinfo{year}{2020}\natexlab{a}.
\newblock \showarticletitle{Sequential recommendation with self-attentive
  multi-adversarial network}. In \bibinfo{booktitle}{\emph{SIGIR 2020}}.
  \bibinfo{pages}{89--98}.
\newblock


\bibitem[\protect\citeauthoryear{Rendle, Freudenthaler, Gantner, and
  Schmidt-Thieme}{Rendle et~al\mbox{.}}{2009}]%
        {rendle2009bpr}
\bibfield{author}{\bibinfo{person}{Steffen Rendle}, \bibinfo{person}{Christoph
  Freudenthaler}, \bibinfo{person}{Zeno Gantner}, {and} \bibinfo{person}{Lars
  Schmidt-Thieme}.} \bibinfo{year}{2009}\natexlab{}.
\newblock \showarticletitle{BPR: Bayesian personalized ranking from implicit
  feedback}. In \bibinfo{booktitle}{\emph{UAI 2009}}.
  \bibinfo{pages}{452--461}.
\newblock


\bibitem[\protect\citeauthoryear{Rendle, Freudenthaler, and
  Schmidt-Thieme}{Rendle et~al\mbox{.}}{2010}]%
        {rendle2010factorizing}
\bibfield{author}{\bibinfo{person}{Steffen Rendle}, \bibinfo{person}{Christoph
  Freudenthaler}, {and} \bibinfo{person}{Lars Schmidt-Thieme}.}
  \bibinfo{year}{2010}\natexlab{}.
\newblock \showarticletitle{Factorizing personalized markov chains for
  next-basket recommendation}. In \bibinfo{booktitle}{\emph{WWW 2010}}.
  \bibinfo{pages}{811--820}.
\newblock


\bibitem[\protect\citeauthoryear{Sahebi and Brusilovsky}{Sahebi and
  Brusilovsky}{2013}]%
        {SahebiB13}
\bibfield{author}{\bibinfo{person}{Shaghayegh Sahebi} {and}
  \bibinfo{person}{Peter Brusilovsky}.} \bibinfo{year}{2013}\natexlab{}.
\newblock \showarticletitle{Cross-domain collaborative recommendation in a
  cold-start context: The impact of user profile size on the quality of
  recommendation}. In \bibinfo{booktitle}{\emph{UMAP 2013}}.
  \bibinfo{pages}{289--295}.
\newblock


\bibitem[\protect\citeauthoryear{Sarwar, Karypis, Konstan, and Riedl}{Sarwar
  et~al\mbox{.}}{2001}]%
        {sarwar2001item}
\bibfield{author}{\bibinfo{person}{Badrul~Munir Sarwar},
  \bibinfo{person}{George Karypis}, \bibinfo{person}{Joseph~A Konstan}, {and}
  \bibinfo{person}{John Riedl}.} \bibinfo{year}{2001}\natexlab{}.
\newblock \showarticletitle{Item-based collaborative filtering recommendation
  algorithms}. In \bibinfo{booktitle}{\emph{WWW 2001}}.
  \bibinfo{pages}{285--295}.
\newblock


\bibitem[\protect\citeauthoryear{Shani, Heckerman, and Brafman}{Shani
  et~al\mbox{.}}{2005}]%
        {shani2005mdp}
\bibfield{author}{\bibinfo{person}{Guy Shani}, \bibinfo{person}{David
  Heckerman}, {and} \bibinfo{person}{Ronen~I Brafman}.}
  \bibinfo{year}{2005}\natexlab{}.
\newblock \showarticletitle{An MDP-based recommender system}.
\newblock \bibinfo{journal}{\emph{J Mach Learn Res 2005}}  \bibinfo{volume}{6}
  (\bibinfo{year}{2005}), \bibinfo{pages}{1265--1295}.
\newblock


\bibitem[\protect\citeauthoryear{Shapira, Rokach, and Freilikhman}{Shapira
  et~al\mbox{.}}{2013}]%
        {ShapiraRF13}
\bibfield{author}{\bibinfo{person}{Bracha Shapira}, \bibinfo{person}{Lior
  Rokach}, {and} \bibinfo{person}{Shirley Freilikhman}.}
  \bibinfo{year}{2013}\natexlab{}.
\newblock \showarticletitle{Facebook single and cross domain data for
  recommendation systems}.
\newblock \bibinfo{journal}{\emph{UMUAI 2013}}  \bibinfo{volume}{23}
  (\bibinfo{year}{2013}), \bibinfo{pages}{211--247}.
\newblock


\bibitem[\protect\citeauthoryear{Song, Xiao, Wang, Charlin, Zhang, and
  Tang}{Song et~al\mbox{.}}{2019}]%
        {song2019session}
\bibfield{author}{\bibinfo{person}{Weiping Song}, \bibinfo{person}{Zhiping
  Xiao}, \bibinfo{person}{Yifan Wang}, \bibinfo{person}{Laurent Charlin},
  \bibinfo{person}{Ming Zhang}, {and} \bibinfo{person}{Jian Tang}.}
  \bibinfo{year}{2019}\natexlab{}.
\newblock \showarticletitle{Session-based social recommendation via dynamic
  graph attention networks}. In \bibinfo{booktitle}{\emph{WSDM 2019}}.
  \bibinfo{pages}{555--563}.
\newblock


\bibitem[\protect\citeauthoryear{Sun, Liu, Wu, Pei, Lin, Ou, and Jiang}{Sun
  et~al\mbox{.}}{2019}]%
        {bert_sr_2019}
\bibfield{author}{\bibinfo{person}{Fei Sun}, \bibinfo{person}{Jun Liu},
  \bibinfo{person}{Jian Wu}, \bibinfo{person}{Changhua Pei},
  \bibinfo{person}{Xiao Lin}, \bibinfo{person}{Wenwu Ou}, {and}
  \bibinfo{person}{Peng Jiang}.} \bibinfo{year}{2019}\natexlab{}.
\newblock \showarticletitle{BERT4Rec: Sequential recommendation with
  bidirectional encoder representations from transformer}. In
  \bibinfo{booktitle}{\emph{CIKM 2019}}. \bibinfo{pages}{1441--1450}.
\newblock


\bibitem[\protect\citeauthoryear{Tan, Xu, and Liu}{Tan et~al\mbox{.}}{2016}]%
        {improveGRU4rec}
\bibfield{author}{\bibinfo{person}{Yong~Kiam Tan}, \bibinfo{person}{Xinxing
  Xu}, {and} \bibinfo{person}{Yong Liu}.} \bibinfo{year}{2016}\natexlab{}.
\newblock \showarticletitle{Improved recurrent neural networks for
  session-based recommendations}. In \bibinfo{booktitle}{\emph{RecSys 2016}}.
  \bibinfo{pages}{17--22}.
\newblock


\bibitem[\protect\citeauthoryear{Tiroshi, Berkovsky, K{\^{a}}afar, Chen, and
  Kuflik}{Tiroshi et~al\mbox{.}}{2013}]%
        {TiroshiBKCK13}
\bibfield{author}{\bibinfo{person}{Amit Tiroshi}, \bibinfo{person}{Shlomo
  Berkovsky}, \bibinfo{person}{Mohamed~Ali K{\^{a}}afar},
  \bibinfo{person}{Terence Chen}, {and} \bibinfo{person}{Tsvi Kuflik}.}
  \bibinfo{year}{2013}\natexlab{}.
\newblock \showarticletitle{Cross social networks interests predictions based
  on graph features}. In \bibinfo{booktitle}{\emph{RecSys 2013}}.
  \bibinfo{pages}{319--322}.
\newblock


\bibitem[\protect\citeauthoryear{Tiroshi and Kuflik}{Tiroshi and
  Kuflik}{2012}]%
        {TiroshiK12}
\bibfield{author}{\bibinfo{person}{Amit Tiroshi} {and} \bibinfo{person}{Tsvi
  Kuflik}.} \bibinfo{year}{2012}\natexlab{}.
\newblock \showarticletitle{Domain ranking for cross domain collaborative
  filtering}. In \bibinfo{booktitle}{\emph{UMAP 2012}}.
  \bibinfo{pages}{328--333}.
\newblock


\bibitem[\protect\citeauthoryear{Wang, Zhang, Wang, Zhao, Li, Xie, and
  Guo}{Wang et~al\mbox{.}}{2018}]%
        {RippleNet}
\bibfield{author}{\bibinfo{person}{Hongwei Wang}, \bibinfo{person}{Fuzheng
  Zhang}, \bibinfo{person}{Jialin Wang}, \bibinfo{person}{Miao Zhao},
  \bibinfo{person}{Wenjie Li}, \bibinfo{person}{Xing Xie}, {and}
  \bibinfo{person}{Minyi Guo}.} \bibinfo{year}{2018}\natexlab{}.
\newblock \showarticletitle{Ripplenet: Propagating user preferences on the
  knowledge graph for recommender systems}. In \bibinfo{booktitle}{\emph{CIKM
  2018}}. \bibinfo{pages}{417--426}.
\newblock


\bibitem[\protect\citeauthoryear{Wang, Zhao, Xie, Li, and Guo}{Wang
  et~al\mbox{.}}{2019d}]%
        {KGCN}
\bibfield{author}{\bibinfo{person}{Hongwei Wang}, \bibinfo{person}{Miao Zhao},
  \bibinfo{person}{Xing Xie}, \bibinfo{person}{Wenjie Li}, {and}
  \bibinfo{person}{Minyi Guo}.} \bibinfo{year}{2019}\natexlab{d}.
\newblock \showarticletitle{Knowledge graph convolutional networks for
  recommender systems}. In \bibinfo{booktitle}{\emph{WWW 2019}}.
  \bibinfo{pages}{3307--3313}.
\newblock


\bibitem[\protect\citeauthoryear{Wang, Ren, Mei, Chen, Ma, and de~Rijke}{Wang
  et~al\mbox{.}}{2019c}]%
        {wang2019collaborative}
\bibfield{author}{\bibinfo{person}{Meirui Wang}, \bibinfo{person}{Pengjie Ren},
  \bibinfo{person}{Lei Mei}, \bibinfo{person}{Zhumin Chen},
  \bibinfo{person}{Jun Ma}, {and} \bibinfo{person}{Maarten de Rijke}.}
  \bibinfo{year}{2019}\natexlab{c}.
\newblock \showarticletitle{A collaborative session-based recommendation
  approach with parallel memory modules}. In \bibinfo{booktitle}{\emph{SIGIR
  2019}}. \bibinfo{pages}{345--354}.
\newblock


\bibitem[\protect\citeauthoryear{Wang, Guo, Lan, Xu, Wan, and Cheng}{Wang
  et~al\mbox{.}}{2015}]%
        {wang2015learning}
\bibfield{author}{\bibinfo{person}{Pengfei Wang}, \bibinfo{person}{Jiafeng
  Guo}, \bibinfo{person}{Yanyan Lan}, \bibinfo{person}{Jun Xu},
  \bibinfo{person}{Shengxian Wan}, {and} \bibinfo{person}{Xueqi Cheng}.}
  \bibinfo{year}{2015}\natexlab{}.
\newblock \showarticletitle{Learning hierarchical representation model for
  nextbasket recommendation}. In \bibinfo{booktitle}{\emph{SIGIR 2015}}.
  \bibinfo{pages}{403--412}.
\newblock


\bibitem[\protect\citeauthoryear{Wang, He, Cao, Liu, and Chua}{Wang
  et~al\mbox{.}}{2019b}]%
        {wang2019kgat}
\bibfield{author}{\bibinfo{person}{Xiang Wang}, \bibinfo{person}{Xiangnan He},
  \bibinfo{person}{Yixin Cao}, \bibinfo{person}{Meng Liu}, {and}
  \bibinfo{person}{Tat-Seng Chua}.} \bibinfo{year}{2019}\natexlab{b}.
\newblock \showarticletitle{KGAT: Knowledge graph attention network for
  recommendation}. In \bibinfo{booktitle}{\emph{SIGKDD 2019}}.
  \bibinfo{pages}{950--958}.
\newblock


\bibitem[\protect\citeauthoryear{Wang, He, Nie, and Chua}{Wang
  et~al\mbox{.}}{2017}]%
        {itemsilkroad2017}
\bibfield{author}{\bibinfo{person}{Xiang Wang}, \bibinfo{person}{Xiangnan He},
  \bibinfo{person}{Liqiang Nie}, {and} \bibinfo{person}{Tat-Seng Chua}.}
  \bibinfo{year}{2017}\natexlab{}.
\newblock \showarticletitle{Item silk road: Recommending items from information
  domains to social users}. In \bibinfo{booktitle}{\emph{SIGIR 2017}}.
  \bibinfo{pages}{185--194}.
\newblock


\bibitem[\protect\citeauthoryear{Wang, Feng, Guo, Chu, and Hwang}{Wang
  et~al\mbox{.}}{2019a}]%
        {wang2019solving}
\bibfield{author}{\bibinfo{person}{Yaqing Wang}, \bibinfo{person}{Chunyan
  Feng}, \bibinfo{person}{Caili Guo}, \bibinfo{person}{Yunfei Chu}, {and}
  \bibinfo{person}{Jenq-Neng Hwang}.} \bibinfo{year}{2019}\natexlab{a}.
\newblock \showarticletitle{Solving the sparsity problem in recommendations via
  cross-domain item embedding based on co-clustering}. In
  \bibinfo{booktitle}{\emph{WSDM 2019}}. \bibinfo{pages}{717--725}.
\newblock


\bibitem[\protect\citeauthoryear{Wu, Tang, Zhu, Wang, Xie, and Tan}{Wu
  et~al\mbox{.}}{2019}]%
        {SRGNN}
\bibfield{author}{\bibinfo{person}{Shu Wu}, \bibinfo{person}{Yuyuan Tang},
  \bibinfo{person}{Yanqiao Zhu}, \bibinfo{person}{Liang Wang},
  \bibinfo{person}{Xing Xie}, {and} \bibinfo{person}{Tieniu Tan}.}
  \bibinfo{year}{2019}\natexlab{}.
\newblock \showarticletitle{Session-based recommendation with graph neural
  networks}. In \bibinfo{booktitle}{\emph{AAAI 2019}}.
  \bibinfo{pages}{346--353}.
\newblock


\bibitem[\protect\citeauthoryear{Wu, Liu, Chen, He, Lv, Cao, and Hu}{Wu
  et~al\mbox{.}}{2013}]%
        {wu2013personalized}
\bibfield{author}{\bibinfo{person}{Xiang Wu}, \bibinfo{person}{Qi Liu},
  \bibinfo{person}{Enhong Chen}, \bibinfo{person}{Liang He},
  \bibinfo{person}{Jingsong Lv}, \bibinfo{person}{Can Cao}, {and}
  \bibinfo{person}{Guoping Hu}.} \bibinfo{year}{2013}\natexlab{}.
\newblock \showarticletitle{Personalized next-song recommendation in online
  karaokes}. In \bibinfo{booktitle}{\emph{RecSys 2013}}.
  \bibinfo{pages}{137--140}.
\newblock


\bibitem[\protect\citeauthoryear{Xian, Fu, Muthukrishnan, de~Melo, and
  Zhang}{Xian et~al\mbox{.}}{2019}]%
        {xian2019reinforcement}
\bibfield{author}{\bibinfo{person}{Yikun Xian}, \bibinfo{person}{Zuohui Fu},
  \bibinfo{person}{S Muthukrishnan}, \bibinfo{person}{Gerard de Melo}, {and}
  \bibinfo{person}{Yongfeng Zhang}.} \bibinfo{year}{2019}\natexlab{}.
\newblock \showarticletitle{Reinforcement knowledge graph reasoning for
  explainable recommendation}. In \bibinfo{booktitle}{\emph{SIGIR 2019}}.
  \bibinfo{pages}{285--294}.
\newblock


\bibitem[\protect\citeauthoryear{Yap, Li, and Philip}{Yap
  et~al\mbox{.}}{2012}]%
        {yap2012effective}
\bibfield{author}{\bibinfo{person}{Ghim-Eng Yap}, \bibinfo{person}{Xiao-Li Li},
  {and} \bibinfo{person}{S~Yu Philip}.} \bibinfo{year}{2012}\natexlab{}.
\newblock \showarticletitle{Effective next-items recommendation via
  personalized sequential pattern mining}. In \bibinfo{booktitle}{\emph{DASFAA
  2012}}. \bibinfo{pages}{48--64}.
\newblock


\bibitem[\protect\citeauthoryear{Zhang, Yuan, Lian, Xie, and Ma}{Zhang
  et~al\mbox{.}}{2016}]%
        {zhang2016collaborative}
\bibfield{author}{\bibinfo{person}{Fuzheng Zhang},
  \bibinfo{person}{Nicholas~Jing Yuan}, \bibinfo{person}{Defu Lian},
  \bibinfo{person}{Xing Xie}, {and} \bibinfo{person}{Wei-Ying Ma}.}
  \bibinfo{year}{2016}\natexlab{}.
\newblock \showarticletitle{Collaborative knowledge base embedding for
  recommender systems}. In \bibinfo{booktitle}{\emph{SIGKDD 2016}}.
  \bibinfo{pages}{353--362}.
\newblock


\bibitem[\protect\citeauthoryear{Zhang, Zhao, Liu, Sheng, Xu, Wang, Liu, and
  Zhou}{Zhang et~al\mbox{.}}{2019}]%
        {zhang2019feature}
\bibfield{author}{\bibinfo{person}{Tingting Zhang}, \bibinfo{person}{Pengpeng
  Zhao}, \bibinfo{person}{Yanchi Liu}, \bibinfo{person}{Victor Sheng},
  \bibinfo{person}{Jiajie Xu}, \bibinfo{person}{Deqing Wang},
  \bibinfo{person}{Guanfeng Liu}, {and} \bibinfo{person}{Xiaofang Zhou}.}
  \bibinfo{year}{2019}\natexlab{}.
\newblock \showarticletitle{Feature-level deeper self-attention network for
  sequential recommendation}. In \bibinfo{booktitle}{\emph{IJCAI 2019}}.
  \bibinfo{pages}{4320--4326}.
\newblock


\bibitem[\protect\citeauthoryear{Zhao, Li, Xiao, Deng, and Sun}{Zhao
  et~al\mbox{.}}{2020}]%
        {zhao2020catn}
\bibfield{author}{\bibinfo{person}{Cheng Zhao}, \bibinfo{person}{Chenliang Li},
  \bibinfo{person}{Rong Xiao}, \bibinfo{person}{Hongbo Deng}, {and}
  \bibinfo{person}{Aixin Sun}.} \bibinfo{year}{2020}\natexlab{}.
\newblock \showarticletitle{CATN: Cross-domain recommendation for cold-start
  users via aspect transfer network}. In \bibinfo{booktitle}{\emph{SIGIR
  2020}}. \bibinfo{pages}{229--238}.
\newblock


\bibitem[\protect\citeauthoryear{Zhao, Wang, He, Wen, Chang, and Li}{Zhao
  et~al\mbox{.}}{2016}]%
        {zhao2016tkdd}
\bibfield{author}{\bibinfo{person}{Wayne~Xin Zhao}, \bibinfo{person}{Jinpeng
  Wang}, \bibinfo{person}{Yulan He}, \bibinfo{person}{Ji-Rong Wen},
  \bibinfo{person}{Edward~Y Chang}, {and} \bibinfo{person}{Xiaoming Li}.}
  \bibinfo{year}{2016}\natexlab{}.
\newblock \showarticletitle{Mining product adopter information from online
  reviews for improving product recommendation}.
\newblock \bibinfo{journal}{\emph{TKDD 2016}}  \bibinfo{volume}{10}
  (\bibinfo{year}{2016}), \bibinfo{pages}{1--23}.
\newblock


\bibitem[\protect\citeauthoryear{Zheng, Guo, Chen, Yu, and Jiang}{Zheng
  et~al\mbox{.}}{2020}]%
        {sentiSR2020}
\bibfield{author}{\bibinfo{person}{Lin Zheng}, \bibinfo{person}{Naicheng Guo},
  \bibinfo{person}{Weihao Chen}, \bibinfo{person}{Jin Yu}, {and}
  \bibinfo{person}{Dazhi Jiang}.} \bibinfo{year}{2020}\natexlab{}.
\newblock \showarticletitle{Sentiment-guided sequential recommendation}. In
  \bibinfo{booktitle}{\emph{SIGIR 2020}}. \bibinfo{pages}{1957--1960}.
\newblock


\bibitem[\protect\citeauthoryear{Zhuang, Zhou, Zhang, Ao, Xie, and He}{Zhuang
  et~al\mbox{.}}{2018}]%
        {cross-domain_novelty-seeking_sr}
\bibfield{author}{\bibinfo{person}{Fuzhen Zhuang}, \bibinfo{person}{Yingmin
  Zhou}, \bibinfo{person}{Fuzheng Zhang}, \bibinfo{person}{Xiang Ao},
  \bibinfo{person}{Xing Xie}, {and} \bibinfo{person}{Qing He}.}
  \bibinfo{year}{2018}\natexlab{}.
\newblock \showarticletitle{Cross-domain novelty seeking trait mining for
  sequential recommendation}.
\newblock \bibinfo{journal}{\emph{arXiv preprint arXiv:1803.01542}}
  (\bibinfo{year}{2018}).
\newblock


\bibitem[\protect\citeauthoryear{Zimdars, Chickering, and Meek}{Zimdars
  et~al\mbox{.}}{2001}]%
        {zimdars2001using}
\bibfield{author}{\bibinfo{person}{Andrew Zimdars},
  \bibinfo{person}{David~Maxwell Chickering}, {and}
  \bibinfo{person}{Christopher Meek}.} \bibinfo{year}{2001}\natexlab{}.
\newblock \showarticletitle{Using temporal data for making recommendations}. In
  \bibinfo{booktitle}{\emph{UAI 2001}}. \bibinfo{pages}{580--588}.
\newblock


\end{thebibliography}

\end{document}